\documentclass[prx,aps,10pt,floatfix,nofootinbib,amsmath,superscriptaddress,twocolumn,notitlepage]{revtex4-2}

\usepackage{amsfonts}
\usepackage{amsmath}
\usepackage{amssymb}
\usepackage{graphicx}
\usepackage{makeidx}
\usepackage[hidelinks]{hyperref}
\usepackage[font={footnotesize},labelfont={bf}]{caption}
\usepackage{setspace}
\usepackage{appendix}
\usepackage{soul}
\usepackage{enumitem}

\usepackage{extarrows}

\usepackage[most]{tcolorbox}
\newtcolorbox{myt}[2][]{%
attach boxed title to top center
           = {yshift=-4pt},
colback      = blue!5!white,
colframe     = blue!75!black,
halign       = flush left,
fonttitle    = \bfseries\sffamily,
colbacktitle = blue!65!black,
title        = #2,#1,
enhanced,
}
\newtcolorbox{myd}[2][]{%
attach boxed title to top center
           = {yshift=-4pt},
colback      = violet!5!white,
colframe     = violet!75!black,
halign       = flush left,
fonttitle    = \bfseries\sffamily,
colbacktitle = violet!65!black,
title        = #2,#1,
enhanced,
}
\newtcolorbox{mye}[2][]{%
attach boxed title to top center
           = {yshift=-4pt},
colback      = purple!5!white,
colframe     = purple!75!black,
halign       = flush left,
fonttitle    = \bfseries\sffamily,
colbacktitle = purple!65!black,
title        = #2,#1,
enhanced,
}

\newtcolorbox{myg}[2][]{%
attach boxed title to top center
           = {yshift=-4pt},
colback      = green!5!white,
colframe     = green!50!black,
halign       = flush left,
fonttitle    = \bfseries\sffamily,
colbacktitle = green!65!black,
title        = #2,#1,
enhanced,
}

\providecommand{\U}[1]{\protect\rule{.1in}{.1in}}

\usepackage{graphics}

\usepackage{amsthm}
\usepackage{color}
\usepackage{dsfont}

\usepackage{mathtools}

\def\P{{\mathbf{P}}}

\def\reg{{\rm reg}}

\def\pure{{\rm Pure}}

\def\cp{{\rm CP}}
\def\cptp{{\rm CPTP}}

\def\stoch{{\rm STOCH}}

\def\J{\mathbf{J}}

\def\1{\mathbf{1}}

\def\>{\rangle}
\def\<{\langle}
\def\id{\mathsf{id}}

\def\mE{\mathcal{E}}

\def\mF{\mathcal{F}}
\def\mN{\mathcal{N}}

\def\mS{\mathcal{S}}
\def\mD{\mathcal{D}}

\def\mV{\mathcal{V}}

\newcommand{\supp}{\operatorname{supp}}

\newcommand{\rl}{\rangle\langle}

\renewcommand{\qedsymbol}{\nobreak \ifvmode \relax \else
\ifdim \lastskip<1.5em \hskip-\lastskip \hskip1.5em plus0em
minus0.5em \fi \nobreak \vrule height0.75em width0.5em
depth0.25em\fi}

\renewcommand{\ge}{\geqslant}
\renewcommand{\le}{\leqslant}
\renewcommand{\geq}{\geqslant}
\renewcommand{\leq}{\leqslant}

\theoremstyle{definition}
\newtheorem{innercustomthm}{Theorem}
\newenvironment{thm}[1]
  {\renewcommand\theinnercustomthm{#1}\innercustomthm}
  {\endinnercustomthm}
\newtheorem{theorem}{Theorem}

\newtheorem{corollary}[theorem]{Corollary}
\newtheorem{lemma}[theorem]{Lemma}

\newtheorem{definition}[theorem]{Definition}

\theoremstyle{remark}
\newtheorem*{remark}{Remark}

\newcommand{\bea}{\begin{eqnarray}}
\newcommand{\eea}{\end{eqnarray}}
\newcommand{\be}{\begin{equation}}
\newcommand{\ee}{\end{equation}}
\newcommand{\ba}{\begin{equation}\begin{aligned}}
\newcommand{\ea}{\end{aligned}\end{equation}}

\newcommand{\epm}{\end{pmatrix}}
\newcommand{\bpm}{\begin{pmatrix}}

\newcommand{\ebm}{\end{bmatrix}}
\newcommand{\bbm}{\begin{bmatrix}}

\newcommand{\bex}{\begin{exercise}}
\newcommand{\eex}{\end{exercise}}

\newcommand{\ben}{\begin{enumerate}}
\newcommand{\een}{\end{enumerate}}

\def\be{\begin{equation}}
\def\ee{\end{equation}}

\newcommand{\rank}{{\rm Rank}}

\newcommand{\mU}{\mathcal{U}}

\newcommand{\md}{\mathfrak{D}}

\newcommand{\mk}{\mathfrak{K}}

\newcommand{\mi}{\mathfrak{I}}

\newcommand{\mR}{\mathcal{R}}
\newcommand{\mM}{\mathcal{M}}

\newcommand{\D}{\mathbb{D}}
\newcommand{\h}{\mathbb{H}}

\newcommand{\lr}{\rangle\langle}

\newcommand{\tr}{{\rm Tr}}

\newcommand{\da}{\downarrow}

\newcommand{\conv}{{\rm Conv}}

\newcommand{\mbb}[1]{\mathbb{#1}}

\def\H{\mathbb{H}}

\newcommand{\bra}[1]{\langle #1|}
\newcommand{\ket}[1]{|#1\rangle}

\newcommand{\eqdef}{\coloneqq}
\newcommand{\mbN}{\mathbb{N}}
\newcommand{\mbR}{\mathbb{R}}

\def\r{\mathbf{r}}
\def\s{\mathbf{s}}
\def\p{\mathbf{p}}
\def\q{\mathbf{q}}

\def\e{\mathbf{e}}
\def\x{\mathbf{x}}

\def\t{\mathbf{t}}
\def\u{\mathbf{u}}
\def\v{\mathbf{v}}
\def\w{\mathbf{w}}
\def\0{\mathbf{0}}
\def\a{\mathbf{a}}
\def\b{\mathbf{b}}

\def\tA{\tilde{A}}
\def\tB{\tilde{B}}

\def\tR{\tilde{R}}

\def\pr{{\rm Pr}}

\def\prob{{\rm Prob}}

\def\bmyd{\begin{myd}{}
\begin{definition}}
\def\emyd{\end{definition}\end{myd}}

\def\bmyl{\begin{myg}{}
\begin{lemma}}
\def\emyl{\end{lemma}\end{myg}}

\def\bmyt{\begin{myt}{}
\begin{theorem}}
\def\emyt{\end{theorem}\end{myt}}

\def\bmyc{\begin{myg}{}
\begin{corollary}}
\def\emyc{\end{corollary}\end{myg}}

\usepackage{subfiles}

\begin{document}	
	\title{Inevitable Negativity: Additivity Commands Negative Quantum Channel Entropy}
\author{Gilad Gour}
\affiliation{Department of Mathematics, Technion - Israel Institute of Technology, Haifa, Israel}
\author{Doyeong Kim}
\affiliation{Department of Mathematics and Statistics and Institute for Quantum Science and Technology, University of Calgary, Calgary, AB T2N 1N4, Canada}
\author{Takla Nateeboon}
\affiliation{Department of Mathematics and Statistics and Institute for Quantum Science and Technology, University of Calgary, Calgary, AB T2N 1N4, Canada}
\author{Guy Shemesh}
\affiliation{Department of Mathematics, Technion - Israel Institute of Technology, Haifa, Israel}
\author{Goni Yoeli}
\affiliation{Department of Mathematics, Technion - Israel Institute of Technology, Haifa, Israel}

\begin{abstract}
    Quantum channels represent a broad spectrum of operations crucial to quantum information theory, encompassing everything from the transmission of quantum information to the manipulation of various resources. In the domain of states, the concept of majorization serves as a fundamental tool for comparing the uncertainty inherent in both classical and quantum systems. This paper establishes a rigorous framework for assessing the uncertainty in both classical and quantum channels. By employing a specific class of superchannels, we introduce and elucidate three distinct approaches to channel majorization: constructive, axiomatic, and operational. Intriguingly, these methodologies converge to a consistent ordering. This convergence not only provides a robust basis for defining entropy functions for channels but also clarifies the interpretation of entropy in this broader context. Most notably, our findings reveal that any viable entropy function for quantum channels must assume negative values, thereby challenging traditional notions of entropy.
\end{abstract}
\maketitle


\section{Introduction}
Majorization is a pivotal concept in linear algebra, extensively covered in the seminal text by Marshall and Olkin \cite{MOA2011}. Its utility spans a variety of fields, including economics, thermodynamics, and quantum information theory, illustrating its broad applicability and fundamental importance. In quantum information, majorization plays a crucial role in the characterization and manipulation of quantum states, as highlighted by Nielsen's majorization theorem~\cite{Nielsen1999,NV2001}. This theorem posits that the interconvertibility among pure bipartite quantum states via local operations and classical communication (LOCC) is governed by a majorization relation between their Schmidt coefficients. Such insights are not only foundational to the understanding of quantum entanglement but also instrumental in applications like entanglement distillation, quantum state discrimination, and quantum key distribution \cite{HHHO2005,Renner2008}. Over the years, the theory of majorization has evolved into an indispensable toolkit in quantum information science \cite{Partovi2011, PRZ2013, HO2013, GMNS+2015, Renes2016, LQ2020, ZLXW+2023}.

The majorization relation between two probability vectors can characterized via three different approaches. Specifically, for two $m$-dimensional probability vectors $\p$ and $\q$, we state that $\p$ \emph{majorizes} $\q$, and denote $\p \succ \q$, if:
																																																																																  
\begin{enumerate}
\item {\it The Constructive Approach:}
$\p$ can be mixed into $\q$ via a convex combination of permutations,
\begin{equation*}
\q = \sum_{j} s_j P_j\p\;,
\end{equation*}
emulating the effect of randomly relabeling
outcomes~\cite{FGG2013}.
Such operations are called \textit{random-permutations}.
\item {\it The Axiomatic Approach:} 
$\p$ can be mixed into $\q$, where here the \textit{mixing operation} is defined with the minimal stipulation that the uniform distribution $\u$ is already maximally mixed, thus, remaining fixed under mixing.
  That is,
$\q = M\p$
 for some stochastic matrix satisfying $M\u = \u$. 
Such operations are termed \textit{doubly-stochastic} or \textit{unital}.
\item {\it The Operational Approach:} 
The odds of winning any game of chance with the distribution $\p$ exceed those with $\q$.
This is formalized as:
\begin{equation*}
\|\p\|_{(k)} \geq \|\q\|_{(k)}  \quad\quad \forall\; k<m\;,
\end{equation*}
where $\|\cdot\|_{(k)}$ is the $k$'th Ky-Fan norm, interpreted operationally as the probability of winning a \textit{$k$-gambling game}, in which a player makes $k$ distinct predictions about the outcome drawn from 
$\p$~\cite{Gour2024}.
\end{enumerate}

Remarkably, these three characterizations of majorization -- constructive, axiomatic, and operational -- have been demonstrated to coincide, as highlighted in the seminal works of Birkhoff \cite{Birkhoff1946} and Hardy, Littlewood and P\'{o}lya \cite{HLP1934}. Moreover, the set of random permutations and the set of doubly-stochastic matrices turn out to coincide, and these operations are collectively termed {\it mixing operations}. This convergence establishes a robust foundation for understanding the concept of uncertainty in systems described with probability vectors. It is worth noting that the recently introduced concept of conditional majorization~\cite{GWBG2022}, developed through the same three distinct approaches, has consistently been shown to result in the same preorder.

Functions that behave monotonically under majorization are known as Schur convex or Schur concave. Axiomatically, we define \textit{entropy} as an additive Schur concave function~\cite{GT2021,LY1999}. It was recently demonstrated that every entropy function is a convex combination of R\'enyi entropies~\cite{MPS+2021}. Note how the combination of monotonicity and additivity \textit{axioms} provides a sufficient framework to \textit{construct} all entropies.

In this paper, we extend the majorization relation from probability vectors and quantum states to classical and quantum channels. This extension, termed channel majorization, evaluates the uncertainty of a channel's output, held by Bob, given a known input by Alice. Within the classical-channel domain, we propose three distinct definitions of channel majorization based on the constructive, axiomatic, and operational approaches discussed above, demonstrate their alignment and thus reinforce the concept of channel majorization. We also establish a standard form for channels appropriate to channel majorization and provide characterizations in terms of sublinear functionals, as well as concrete characterizations in lower dimensions.
																																																																																																																																																																																										
We proceed to define channel entropy functions as additive monotones of channel majorization. Under these foundational assumptions, we show that the every quantum channel entropy must assume negative values on some quantum channels, aligning with similar findings for conditional entropies in earlier work~\cite{GWBG2022}. Furthermore, we show that monotones that are only weakly additive---additive when taking tensor products of any number of copies of the same channel---do not necessarily assume negative values. This distinction underscores the principle that ``quantum additivity implies negativity''.
																																																																						
\section{Classical Channel Majorization}
Consider four classical systems, $X,X',Y,Y'$, and two classical channels, $\mathcal{N}^{X \to Y}$ and $\mathcal{M}^{X' \to Y}$. Our goal is to understand when $\mN$ can be considered more predictable than $\mM$, that is, to define the relation ``$\mN$ majorizes $\mM$''. Initially, we aim to establish a majorization relation between them under the assumption that the output system $Y$ is identical for both channels, as channel majorization fundamentally compares the uncertainty of two channels relative to the same output system. Therefore, we only consider the case in which $Y=Y'$ through this section. The general case of arbitrary system $Y'$ will be considered later in Sec.~\ref{subsection:different-output-dimension}.

To achieve this, we motivate our definition of channel majorization through the same three distinct approaches that we have previously introduced. As with the method of defining vector majorization, the operational definition frames the channel as a resource in gambling games, whereas, in the constructive and axiomatic approaches, the majorized channel $\mM$ is mixed from $\mN$ via some operation. This mixing operation is a physical process of classical channels, i.e., a classical \textit{superchannel}. The axiomatic approach establishes a minimalistic set of axioms that mixing superchannels must satisfy, while the constructive approach formulates them directly. While these approaches offer different perspectives on how to define channel majorization, we will see that the outcomes of these three approaches ultimately converge.

Throughout this section, we will abbreviate $m\coloneqq |X|, m' \coloneqq |X'|$ and $n\coloneqq |Y|$. Observe that when $m=m'=1$, the channels $\mN$ and $\mM$ can be viewed as probability vectors. Note that any reasonable definition of channel majorization should reduce to vector majorization for this case. We will also write $[n]$ to denote the set of all positive integers from $1$ through $n$.
  
\subsection{Constructive Approach}
In the constructive approach, we propose to construct a mixing operation as it is intuitively suggested.
Specifically, a mixing operation $\Theta$ is a stochastic map obtained by Bob applying a random permutation post-processing (i.e., doubly stochastic map) to his system $Y$ conditioned on information received from Alice's pre-processing channel. We call such superchannels \emph{random permutation superchannels}. Mathematically, for every classical channel $\mN^{X \to Y}$
    \be\label{eq:theta-constructive}
    \Theta\left[\mN\right]\eqdef\mD^{YZ\to Y}\circ\mN^{X\to Y}\circ\mS^{X'\to XZ}\;,
    \ee
where $Z$ is the classical system Alice sends to Bob after she processes her input system $X'$ via the channel $\mS^{X'\to XZ}$. Upon receiving the value $z$ of the classical system $Z$, Bob applies a mixing operation to his system $Y$ described by the doubly stochastic channel 
\be
\mD_z^{Y\to Y}\left(\rho^{Y}\right)\eqdef\mD^{YZ\to Y}\left(\rho^{Y}\otimes|z\lr z|^Z\right)\;,
\ee
for all classical (i.e., diagonal) density matrices $\rho^{Y}$. If $\mD_z$ is doubly stochastic for all $z$, we say that $\mD^{YZ\to Y}$ is a conditionally unital channel.

\bmyd\label{def:majo-constructive}
The channel $\mN$ majorizes $\mM$ \emph{constructively} if there exists a random permutation superchannel $\Theta$ of the form~\eqref{eq:theta-constructive} such that $\mM=\Theta\left[\mN\right]$. 
\emyd
\begin{figure}
    \centering
    \includegraphics[width=\columnwidth]{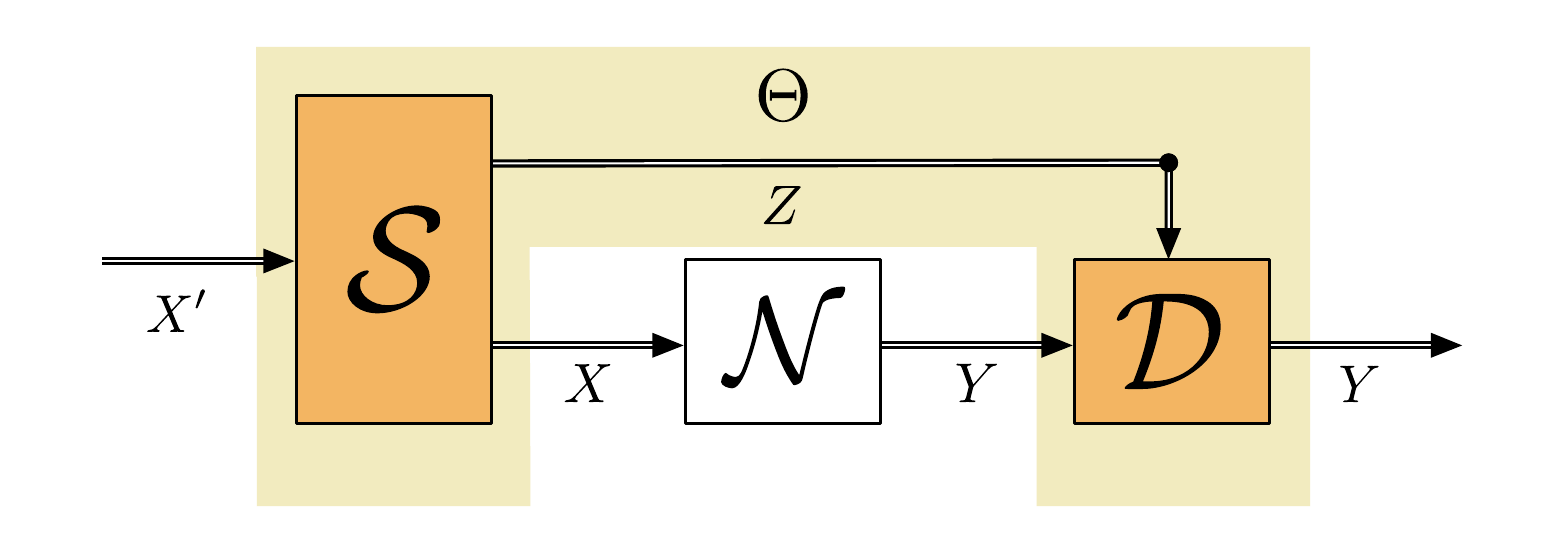}
    \caption{A diagram of random permutation superchannel $\Theta$. The classical preprocessing $\mS$ sends $z$ to Bob. Bob applies a random permutation channel $\mD_z$ corresponding to the received $z$.}
    \label{fig:Constructive-approach}
\end{figure}

\subsection{Axiomatic Approach} 
In the axiomatic approach, we establish our framework based on a minimalistic axiom, which we consider fundamental for any reasonable definition of a mixing operation. Consider, for example, the completely randomizing channel $\mR^{X\to Y}$, which outputs the uniform distribution $\u^{Y}$ for every input state. We refer to this channel as a \emph{uniform channel}, as it consistently outputs $y\in[n]$ with uniform probability $1/n$, regardless of the input $x\in[m]$. Consequently, having access to $X$ does not reduce the uncertainty associated with $Y$, marking this channel as having the highest degree of uncertainty or the least degree of predictability about the output system $Y$.

A mixing superchannel should not decrease the uncertainty of channels and, therefore, must at least preserve the maximality of the uniform channel. 
Specifically, such a superchannel $\Theta$ must satisfy:
\be\label{eq} 
\Theta\left[\mR^{X\to Y}\right]=\mR^{X'\to Y}\;.
\ee
We term such superchannels \emph{uniformity-preserving}. This assumption is notably minimal, as it merely asserts that if $Y$ is initially maximally uncertain, then a mixing operation should not reduce this maximal uncertainty.

Is this the only requirement we should impose on $\Theta$? We propose an additional necessary condition for $\Theta$ to be considered a valid mixing operation, which is deeply rooted in the physical properties similar to the requirement that quantum channels must be completely positive, not just positive. Consider a bipartite channel $\mN^{XZ_0\to YZ_1}$ to be \emph{marginally uniform} if, for any classical (i.e., diagonal) states $\rho^{XZ_0}$, it holds that:
\be
\mN^{XZ_0\to YZ_1}(\rho^{XZ_0})=\u^{Y}\otimes\sigma^{Z_1}
\ee
where $\sigma^{Z_1} \eqdef \mN^{XZ_0\to Z_1}(\rho^{XZ_0})$, with $\mN^{XZ_0\to Z_1} \eqdef \tr_{Y}\circ \mN^{XZ_0\to YZ_1}$, is a classical density matrix, potentially dependent on $\rho^{XZ_0}$. Each marginally uniform channel can be expressed as:
\begin{equation}\label{7}
\mN^{XZ_0\to YZ_1}=\u^{Y}\otimes\mN^{XZ_0\to Z_1}\;,
\end{equation}
where $\mN^{XZ_0\to Z_1}$ is a channel from systems $XZ_0$ to $Z_1$. In this context, a superchannel $\Theta$ must be \emph{completely uniformity-preserving}, meaning for every two systems $Z_0, Z_1$, and any marginally uniform channel $\mN^{XZ_0\to YZ_1}$, the channel
\be\label{m}
\mM^{X'Z_0\to YZ_1}\eqdef\Theta\otimes\mathds{1}^{(Z_0 \to Z_1)}[\mN^{XZ_0\to YZ_1}]\;,
\ee
should also remain marginally uniform. Here, $\mathds{1}^{(Z_0 \to Z_1)}$ denotes the identity superchannel that acts trivially on superoperators of the form $\mF^{Z_0 \to Z_1}$. Interestingly, we demonstrate in the appendix that there are uniformity-preserving superchannels that are not \emph{completely} uniformity-preserving. Therefore, in our axiomatic approach, it is insufficient for $\Theta$ to merely preserve uniformity; it must ensure complete uniformity preservation. Thus, we arrive at the following axiomatic definition of channel-majorization:

\bmyd\label{def}
$\mN^{X\to Y}$ majorizes $\mM^{X'\to Y}$ \emph{axiomatically} if there exists a completely uniformity-preserving superchannel $\Theta$ such that $\mM=\Theta[\mN]$.
\emyd

\begin{figure}[h]\centering    
    \includegraphics[width=\columnwidth]{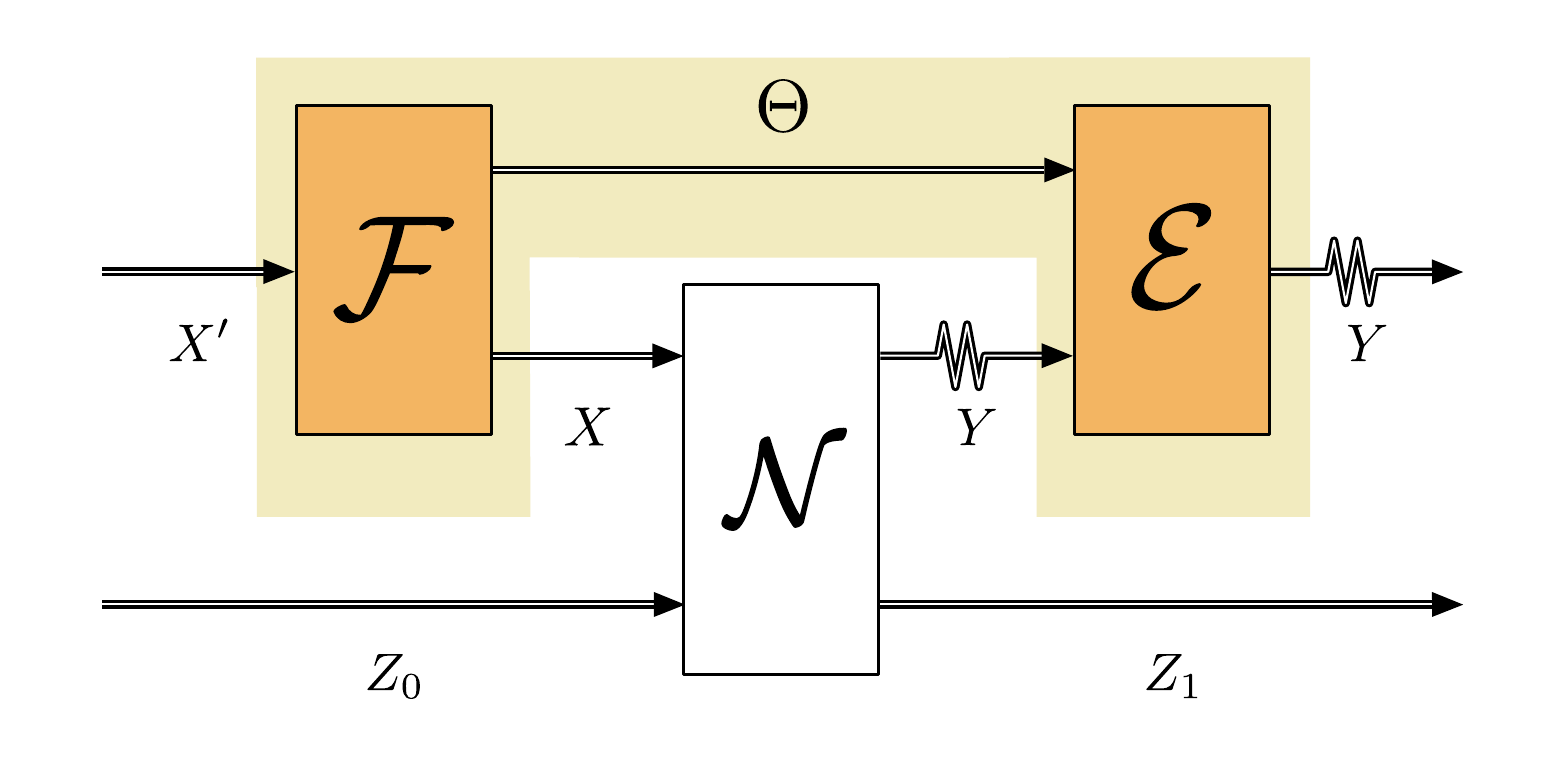}
  \caption{\small A completely uniformity-preserving superchannel. Sharply squiggly lines represent systems in their maximally mixed (i.e., uniform) state.}
  \label{fig:cup}
\end{figure} 

\subsection{Operational approach}

\begin{figure}
    \centering
    \includegraphics[width = \columnwidth]{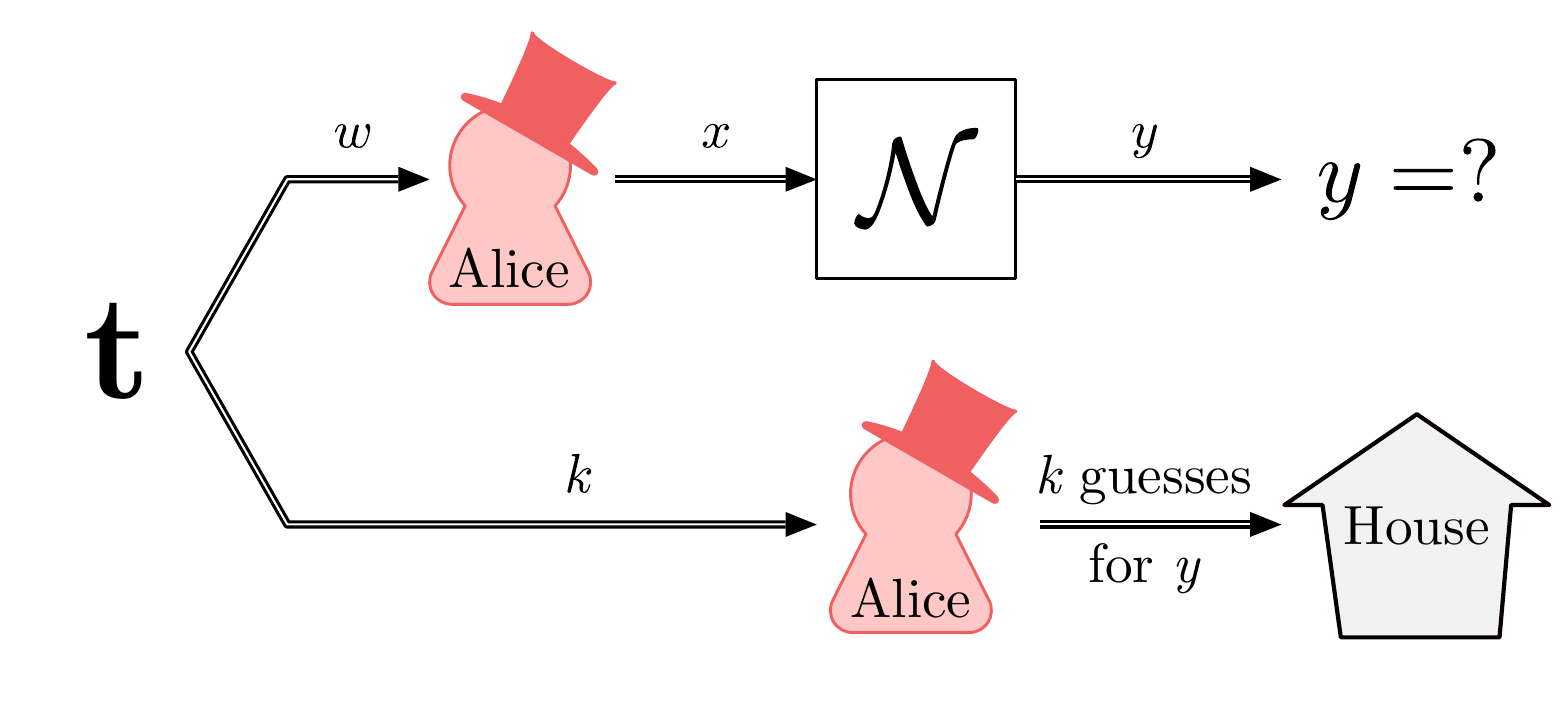}
    \caption{A diagram of a $\t$-gambling game with channel $N^{X \to Y}$. The $\t$ source determines which $k$-game to play. Initially, the player, Alice, learns $k$ partially from $w$, based on which she chooses $x$ to input the channel $\mN$. Once the $k$ value is announced, she provides $k$ guesses of the value $y$ based on her chosen $x$.}
    \label{fig:Diagram_Game-of-chance}
\end{figure}

In the operational approach, we explore a gambling game utilizing a given channel $\mN^{X\to Y}$, where $m \eqdef |X|$ and $n \eqdef |Y|$. This game, previously discussed in \cite{BGG2022}, allows a player complete control over the input system $X$ and challenges them to accurately predict the output $Y$. To illustrate the core concept, we begin with a simplified version of the game, similar to the $k$-game introduced earlier, where the player makes $k$ predictions about $Y$, the output of channel $\mN$, after selecting the optimal input for $\mN$. The maximal winning probability for this $k$-game is given as:
\begin{equation}
    \pr_k(\mN) = \max_{x\in[m]} \Vert\mN(|x\lr x|)\Vert_{(k)},
\end{equation}
where the maximization covers all channel inputs. Each $\mN(|x\rangle \langle x|)$ corresponds to a diagonal density matrix, effectively represented by a probability vector $\p_x \in \prob(n)$. Note that the components of $\p_x$, denoted as $\{p_{y| x}\}_{y\in[n]}$, representing the transition matrix of the classical channel $\mN$. With these notations, the maximum winning probability is simplified to 
\begin{equation}
    \pr_k(\mN) = \max_{x\in[m]} \Vert \p_x\Vert_{(k)}
\end{equation} 

This $k$-game represents just one of many potential gambling games that can be conducted with a classical channel. We aim to generalize this concept and introduce the most comprehensive gambling game achievable with a single use of the channel $\mN$. Consider a scenario where $k\in[n]$ is chosen randomly according to a distribution $\t \in \prob(n)$. In this setup, the player initially knows only the distribution $\t$, not $k$ itself. The player may learn $k$ either before or after selecting the channel input, leading to two distinct optimal winning probabilities:
\begin{align}
    &\text{learn before:}\quad \sum_{k \in[n]} t_k \max _{x \in[m]}\left\|\mathbf{p}_x\right\|_{(k)} \\
    &\text{learn after: } \quad \max _{x \in[m]} \sum_{k \in[n]} t_k\left\|\mathbf{p}_x\right\|_{(k)} \text {, }
\end{align}
That is, when the player is aware of the specific $k$-game to be played, she selects the input $x \in [m]$ that maximizes her winning probability $\|\p_x\|_{(k)}$. However, if she does not know $k$, her optimal strategy involves selecting $x$ to maximize the average probability $\sum_{k \in[n]} t_k\left\|\mathbf{p}_x\right\|_{(k)}$.

The two game variations can be unified into a family of games, which we call the \emph{$\t$-gambling game}. Unlike the discussion above, here we assume that $\t\in \prob(n\ell)$ is a joint probability vector whose components are denoted by $t_{kw}$, where $k\in[n]$ signifies the selection of the $k$-game to be played, and $w\in[\ell]$ signifies the information the player acquires before choosing the channel's input. We also denote by $t_w \eqdef\sum_{k\in[n]} t_{kw}$ the probability that a player will receive the value $w$ and by $t_{k|w}\eqdef t_{kw}/t_w$ the conditional probability that a $k$-game will be played given that the player received $w$.
If there is no correlation between the systems, i.e., $t_{k\vert w}$ does not depend on $w$, then the game reduces to the game in which the player learns $k$ after selecting the input to the channel. On the other hand, the case of perfect correlation between the systems, i.e., $t_{k\vert w} = \delta_{kw}$, corresponds to the case where the player knows $k$ before the input selection. 

Once the player receives $w\in [\ell]$, the player chooses an optimal choice $x\in[m]$ that maximizes $\sum_{k\in[n]} t_{k\vert w} \Vert \p_x\Vert _{(k)}$ as the channel input. Thus, the optimal winning probability for such a $\t$-game with a channel $\mN$ can be expressed as: 
\begin{align}\label{eq:prob-win-T-game}	 
    \pr_\t(\mN) & =\sum_{w \in[\ell]} t_{w} \max _{x \in[m]} \sum_{k \in[n]} t_{k\vert w}\left\|\p_x\right\|_{(k)}\;.
\end{align}
The $\t$-game above represents the most general gambling game that can be played with a single use of a channel.
\bmyd
	   
We say that $\mN$ majorizes $\mM$ \emph{operationally} if for every $\t$-game
    \begin{equation}
        \pr_\t(\mN)\geq \pr_\t(\mM).
    \end{equation}
\emyd

\subsection{Equivalence}
The three distinct approaches proposed above lead to the same preorder on classical channels.
\bmyt\label{th:equivmaj}
    Given two classical channels $\mN$ and $\mM$, the following are equivalent:
    \ben
    \item $\mN$ majorizes $\mM$ constructively
    \item $\mN$ majorizes $\mM$ axiomatically
    \item $\mN$ majorizes $\mM$ operationally
    \een
\emyt
Henceforth we will simply say that $\mN$ majorizes $\mM$, and denote $\mN \succ \mM$, if any of the equivalent conditions above hold. If $\mN \succ \mM$ and $\mM \succ \mN$ we say that $\mN$ and $\mM$ are equivalent and denote $\mN \sim \mM$.

In proving the equivalence between the constructive and axiomatic approaches, we demonstrate something slightly stronger than what is stated in the theorem. Specifically, we show that a superchannel is completely uniformity-preserving if and only if it is a random permutation (cf. Thm. 3 in \cite{Gour2019}). This means that the constructive and axiomatic approaches produce not only the same concept of channel majorization but also the same concept of mixing. We will therefore refer to both random permutation superchannels and completely uniformity-preserving operations simply as \emph{mixing} operations.

\subsection{The Standard Form}\label{section:standard-form}

Every preorder becomes a partial order when restricted to a set of representatives of equivalent classes. For the majorization preorder on $\prob(n)$, one often chooses $\prob^\da(n)$, the subset of those probability vectors with descending entries. This set is indeed a set of representatives as every $\p \in \prob(n)$ is equivalent (via a permutation) to a unique vector in $\prob^\da (n)$, its \textit{standard form} $\p^\da$.

Here we establish a similar result for channel majorization by introducing a similar standard form. In the definition below we consider a channel $\mN^{X\to Y}$ with $m\eqdef|X|$, $n\eqdef|Y|$ and an $n\times m$ transition (stochastic) matrix $N\eqdef[\p_1\cdots\p_m]$, where for each $x\in[m]$ the probability vector $\p_x$ is the diagonal of $\mN(|x\lr x|)$. 

\bmyd
A channel $\mN$ is given in its \emph{standard form} (with respect to channel majorization) if:
\ben
\item The entries of each of its columns are in descending order.
\item The columns of $\mN$ are in decreasing lexicographic order: For all $x\in[m-1]$ there exists $\ell\in[n]$ such that $\|\p_{x}\|_{(k)}= \|\p_{x+1}\|_{(k)}$ for all $k\in[\ell-1]$ and $\|\p_{x}\|_{(\ell)}> \|\p_{x+1}\|_{(\ell)}$.
\item Every column of $N$ is \emph{not} majorized by a convex combination of other columns of $N$.
\een
\emyd

From this definition it follows that for every channel $\mN^{X\to Y}$ with an $n\times m$ transition matrix $N$ can be transformed into its a standard form with an $n\times m'$ transition matrix $N'$ via the following three steps:
\ben
\item Write the entries in each column of $N$ in descending order.

\item Rearrange the columns lexicographically.
\item Remove every column of $N$ that is majorized by a convex combination of other columns of $N$. This will result in an $n\times m'$ column stochastic matrix $\tilde{N}$ with $m'\leq m$.
\een 
After the third step, the resulting matrix $N'$ is in the standard form and we have
$\mN\sim\mN'$.
			   
As an example, consider the case $n\eqdef|Y|=2$. Here, the transition matrix of a channel $\mN^{X\to Y}$ has a transition matrix of the form
\be\label{sfn2}
N=\bbm
r_1 & r_2 & \cdots & r_m\\
1-r_1 & 1-r_2 & \cdots & 1-r_m
\ebm
\ee					  
where without loss of generality we assume that $1\geq r_1\geq\cdots\geq r_m\geq1/2$. This means that the first column $(r_1,1-r_1)^T$ majorizes all the other columns!
Hence, the standard form of this transition matrix is the channel with trivial input given by the vector (diagonal density matrix)  $(r_1, 1-r_1)^T$. Note that this example demonstrates that if the transition matrix of the standard form of a classical channel $\mN^{X\to Y}$ has more than one column then necessarily $|Y|>2$.

In the next theorem (proven in the appendix) we show that there is no freedom left in the definition above in the sense that if $\mN\sim\mM$, and if both $\mN$ and $\mM$ are given in their standard form, then necessarily $\mN=\mM$.

\bmyt
\label{th:standardform}
Using the above notations, $\mN^{X\to Y} \sim \mM^{X'\to Y}$ if and only if $\mN^{X\to Y}=\mM^{X'\to Y}$ (in particular, $X \cong X'$).
\emyt
\subsection{Characterization}
A classical channel can be fully represented with a transition matrix. One can look at the transition matrix as a collection of probability vectors that are conditioned on the input. From this perspective, we can relate channel majorization to the majorization relation between sets of probability vectors. Compactly, classical channel majorization can be characterized by set majorization, which is defined as follows.
\bmyd
    \label{set majorization}
        Let $\mk_1,\mk_2\subseteq\prob\left(n\right)$. We say that $\mk_1\succ \mk_2$ (as sets) if for every $\mathbf{q}\in \mk_2$ there exists $\textbf{p}\in \mk_1$ such that $\textbf{p}\succ\textbf{q}$.
\emyd
In the following theorem we consider two classical channels $\mN^{X\to Y}$ and $\mM^{X'\to Y}$, with $n\eqdef|Y|$, $m\eqdef|X|$, and $m'\eqdef|X'|$. Moreover, we denote each column of the transition matrices of $\mN$ and $\mM$ as
\be
    \p_{x}\eqdef\mN\left(|x\lr x|\right)\quad\quad\forall\;x\in[m]
\ee
and
\be\label{017}
    \q_{y}\eqdef\mM\left(|y\lr y|\right)\quad\quad\forall\;y\in[m']\;.
\ee
Recall that in the classical domain, we view diagonal density matrices as probability vectors. Since we assume that the transition matrices of $\mN$ and $\mM$ are given in their standard form, we have in particular that $\p_x=\p_x^\da$ for all $x\in[m]$, and similarly $\q_{y}=\q_y^\da$ for all $y\in[m']$.
Finally, we will denote by $\conv(\mN)$ and $\conv(\mM)$ the convex hulls of the sets $\{\p_1,\ldots\p_{m}\}$ and $\{\q_1,\ldots\q_{m'}\}$, respectively. In the following theorem, we provide several characterizations of channel majorization. In the appendix, we also show that whether $\mN$ majorizes $\mM$ can be determined efficiently with a linear program.

\bmyt\label{th:tfae-characterization}
The following are equivalent:
\ben
\item $\mN\succ\mM$.
\item $\conv(\mN) \succ \conv(\mM)$.
\item There exists an $m\times m'$ stochastic matrix $S=(s_{x|w})$ such that
\be\label{eq:spsq}
\sum_{x\in[m]}s_{x|w}\p_{x}\succ \q_{w}\quad\quad\forall\;w\in[m']\;.
\ee
\item For all $\s\in\prob^\da(n)$
\be
\max_{x\in[m]}\s\cdot\p_{x} \geq \max_{w\in[m']}\s\cdot\q_{w}\;.
\ee
\een
\emyt
\subsection{Lower-dimensional Cases}\label{sec:g}
In the simple scenario where $|X| = |X'| = 1$, it is straightforward to verify that the majorization relationship $\mathcal{N} \succ \mathcal{M}$ corresponds to vector majorization. Similarly, when dealing with a two-dimensional output space, i.e., $|Y| = 2$, the situation remains straightforward as each channel $\mathcal{N}^{X \to Y}$ can be represented in a standard form that assumes a trivial input, as outlined in the discussion around Eq.~\eqref{sfn2}. We now turn our attention to more complex cases. Again we denote $m \eqdef |X|$, $m' \eqdef |X'|$, and $n \eqdef |Y|$, assuming that the channels are described in their standard forms to facilitate a more nuanced exploration of their properties.
 
\noindent \textbf{Example 1.} The case $m=1$ (and $n,m'>1$). In this case $\mN^{X\to Y}$ can be viewed as a probability vector $\p\in\prob(n)$. Using the notations in~\eqref{017} we obtain from~\eqref{eq:spsq} that $\p\succ\mM$ if and only if $\p\succ\q_w$ for all $w\in[m']$. It is worth mentioning that it is well known~\cite{Bapat1991,BBH+2019,YXH+2023} that the set $\{\q_w\}_{w\in[m']}$ has an optimal upper bound $\r\in\prob(n)$ with the property that $\r\succ\q_w$ for all $w\in[m']$, and if also $\p\succ\q_w$ for all $w\in[m']$ then necessarily $\p\succ\r$. We give the explicit form of $\r$ in the appendix and conclude here that $\p\succ\mM$ if and only if $\p\succ\r$.

\noindent \textbf{Example 2.} The case $m=2$.
Interestingly, this case can also be solved analytically.
In this case, the matrix $S$ that appears in~\eqref{eq:spsq} has only two rows, and since it is column stochastic, each of its columns has the form $(t_w, 1-t_w)^T$, where for each $w\in[m']$, $t_w\eqdef s_{1|w}$. Hence, we get from the theorem above, particularly~\eqref{eq:spsq}, that $\mN\succ\mM$ if and only if for every $w\in[m']$ there exists $t_{w}\in[0,1]$ such that
\be\label{0}
t_{w}\p_1+(1-t_{w})\p_2\succ \q_{w}\;.
\ee
Since the channels are given in their standard form, the components of all the probability vectors above are arranged in non-increasing order. Thus,
in terms of the Ky-Fan norms, the relation above is equivalent to the condition that for all $w\in[m']$ and all $k\in[n]$ we have
\be
t_{w}\|\p_1\|_{(k)}+(1-t_{w})\|\p_2\|_{(k)}\geq \|\q_{w}\|_{(k)}\;,
\ee
or equivalently,   
\be\label{divide}
t_{w}\left(\|\p_1\|_{(k)}-\|\p_2\|_{(k)}\right)\geq \|\q_{w}\|_{(k)}-\|\p_2\|_{(k)}\;.
\ee
Let $\mi_+$, $\mi_-$, and $\mi_0$ denote the subsets of integers $k\in[n]$ for which $\|\p_1\|_{(k)}-\|\p_2\|_{(k)}$ is positive, negative, or zero, respectively. Eq. \eqref{divide} leads to the following conditions:
\begin{enumerate}
    \item 
        $\quad\begin{aligned}
        t_w\geq \mu_w\eqdef\max_{k\in\mi_+}\frac{\|\q_{w}\|_{(k)}-\|\p_{2}\|_{(k)}}{\|\p_{1}\|_{(k)}-\|\p_{2}\|_{(k)}}\;.
    \end{aligned}$
    \item $\quad
    \begin{aligned}
        t_w\leq\nu_{w}\eqdef\min_{k\in\mi_-}\frac{\|\p_{2}\|_{(k)}-\|\q_{w}\|_{(k)}}{\|\p_{2}\|_{(k)}-\|\p_{1}\|_{(k)}}\;.
    \end{aligned}$
    \item $\quad\begin{aligned}
        \|\p_2\|_{(k)}\geq \|\q\|_{(k)}\;, && \forall k \in \mi_0\;.
    \end{aligned}$
\end{enumerate}
Hence, we obtain that $\mN\succ\q$ if and only if the following conditions hold:
\ben
\item $\nu_w\geq\mu_w$.
\item For all $k\in\mi_0$, $\|\p_2\|_{(k)}\geq\|\q_{w}\|_{(k)}$.
\een
These two conditions ensure that for each $w\in[m']$ there exists $t_w\in[0,1]$ satisfying~\eqref{0}. We conclude that $\mN\succ \mM$ if and only if the above conditions hold for any column $\q$ of $\mM$.

\section{Quantum Channel Majorization}\label{sec:Quantum-channel-majorization}
In this section, we extend the definition of channel majorization preorder to the quantum domain. The primary objective is to formulate a robust and precise definition of majorization between quantum channels that will effectively capture the essence of uncertainty or the degree of predictability associated with them.
Since the extension of games of chance does not follow trivially~\cite{BGWG2021}, our definition of channel majorization will be based only on two of the three approaches: the axiomatic and the constructive. As before, although each of the two approaches defines mixing operations from its own perspective, they will eventually prove to be equivalent characterizations of the same superchannel and hence, induce the same preorder of quantum channel majorization.

Our focus will be on examining the transition from one quantum channel $\mathcal{N}\in\cptp(A\to B)$ to another quantum channel $\mathcal{M}\in\cptp(A'\to B)$. The superchannel facilitating this evolution will be denoted by $\Theta$. It's important to note that we consider the conversion of the dynamics $A\to B$ into the dynamics $A'\to B$; in essence, it represents the evolution of evolution itself. 

Every superchannel $\Theta$ can be implemented with an auxiliary system $R$, a channel $\mE\in\cptp(RB\to B)$, and  isometry $\mV\in\cptp(A'\to RA)$. For any channel $\mN \in \cptp(A \to B)$, the action of $\Theta$ is described by~\cite{CDP2008,Gour2019}
\be\label{theta}
\Theta\left[\mN^{A\to B}\right]=\mE^{RB\to B}\circ\mN^{A\to B}\circ\mV^{A'\to RA}\;.
\ee
Furthermore, as established in~\cite{Gour2019}, the minimal dimension of $R$ is determined by the rank of the marginal of the Choi matrix $\J_\Theta^{AA'}$, which implies that $|R| \leq |AA'|$. A superchannel $\Theta$ is said to be in its \emph{standard form} if it is represented as in the above equation with $|R|$ minimized to its feasible lowest dimension. Note that the above superchannel can be interpreted as Alice applies the pre-processing isometry $\mV^{A'\to RA}$ and sends the (correlated) quantum system $R$ to Bob, who then applies the post-processing channel $\mE^{RB\to B}$. 

Quantum channel majorization is a preorder relationship defined between two quantum channels, $\mathcal{N} \in \mathrm{CPTP}(A \to B)$ and $\mathcal{M} \in \mathrm{CPTP}(A' \to B)$. Analogous to its classical counterpart, this preorder is established through the concept of a mixing superchannel, denoted as $\Theta$. The primary objective is to formulate a robust and precise definition of such a mixing superchannel, which effectively captures the essence of uncertainty or the degree of predictability in quantum channels.

\subsection{Mixing Superchannels}
\subsubsection*{The Constructive Approach}
Consider the superchannel $\Theta$ as given in~\eqref{theta}. This form of a superchannel is the same one we used in the classical domain, but with all quantum systems (e.g., $A$, $A'$, etc) replaced with classical counterparts (cf.~\eqref{eq:theta-constructive}). Building on this similarity, recall that in the classical domain, a classical mixing superchannel was defined in such a way that upon receiving the classical system $R$ (which we denoted earlier by $Z$), Bob applies a doubly stochastic channel. That is, we required that the channel $\mE_z^{B\to B}(\cdot)\eqdef\mE^{RB\to B}(|z\lr z|\otimes(\cdot))$ is doubly stochastic. 

Similarly, in the quantum case, we replace the classical message $z$ that was received from Alice with a quantum state $\tau^R$. That is, we require that Bob processes the information sent by Alice on the system $R$, which arrives in the form of a marginal quantum state $\tau^R$, and chooses a doubly stochastic channel 
\be\label{r}
\mE_\tau^{B \to B}(\omega^{B}):=\mE^{RB \to B} (\tau^{R}\otimes\omega^{B})\;.
\ee
That is, we require that $\mE_\tau$ is a doubly stochastic channel for every choice of $\tau\in\md(R)$.
Every channel $\mE^{RB\to B}$ that satisfies the condition above is said to be \emph{conditionally unital}.

\bmyd
    Let $\Theta$ be in its standard form as in~\eqref{theta}. $\Theta$ is said to be a \emph{mixing} operation if 
     $\mE^{RB \to B}$ is conditionally unital.
\emyd

\subsubsection*{The Axiomatic Approach}
We begin by extending the concept of marginal uniformity to the quantum domain. Let $\mN\in\cptp(AC_0\to BC_1)$ and denote its marginal
\be
\mN^{AC_0\to C_1}\eqdef\tr_{B}\circ\mN^{AC_0\to BC_1}\;.
\ee
Similar to the classical case, we say that a quantum channel
$\mN\in\cptp(AC_0\to BC_1)$ is \emph{marginally} uniform with respect to $B$ if 
\begin{equation}
    \mN^{AC_0\to BC_1} = \u^{B} \otimes \mN^{AC_0 \to C_1}\;.
\end{equation}
Equivalently, $\mN^{AC_0\to BC_1}$ is marginally uniform if
\be
\mN^{AC_0\to BC_1} =\mR^{B}\circ\mN^{AC_0\to BC_1}\;,
\ee 
where $\mR^{B}\in\cptp(B\to B)$ is the uniform channel (also known as a completely randomizing channel or completely depolarizing channel). Marginal uniform channels generate maximal uncertainty in system $B$, regardless of the information in $A$. We expect this maximality to be preserved under mixing operations. This leads to the following definition:

\bmyd
A superchannel $\Theta$ is said to be \emph{completely uniformity-preserving} if for every pair of systems $C_0$ and $C_1$, and every marginally uniform channel $\mN^{AC_0\to BC_1}$ with respect to $B$, the channel
\begin{equation*}
\mM^{A'C_0\to B'C_1}\eqdef\Theta\otimes \mathds{1}^{(C_0 \to C_1)}
\left[\mN^{AC_0\to BC_1}\right]
\end{equation*}
is marginally uniform with respect to $B'$. 
\emyd

\begin{figure}\centering    
    \includegraphics[width=\columnwidth]{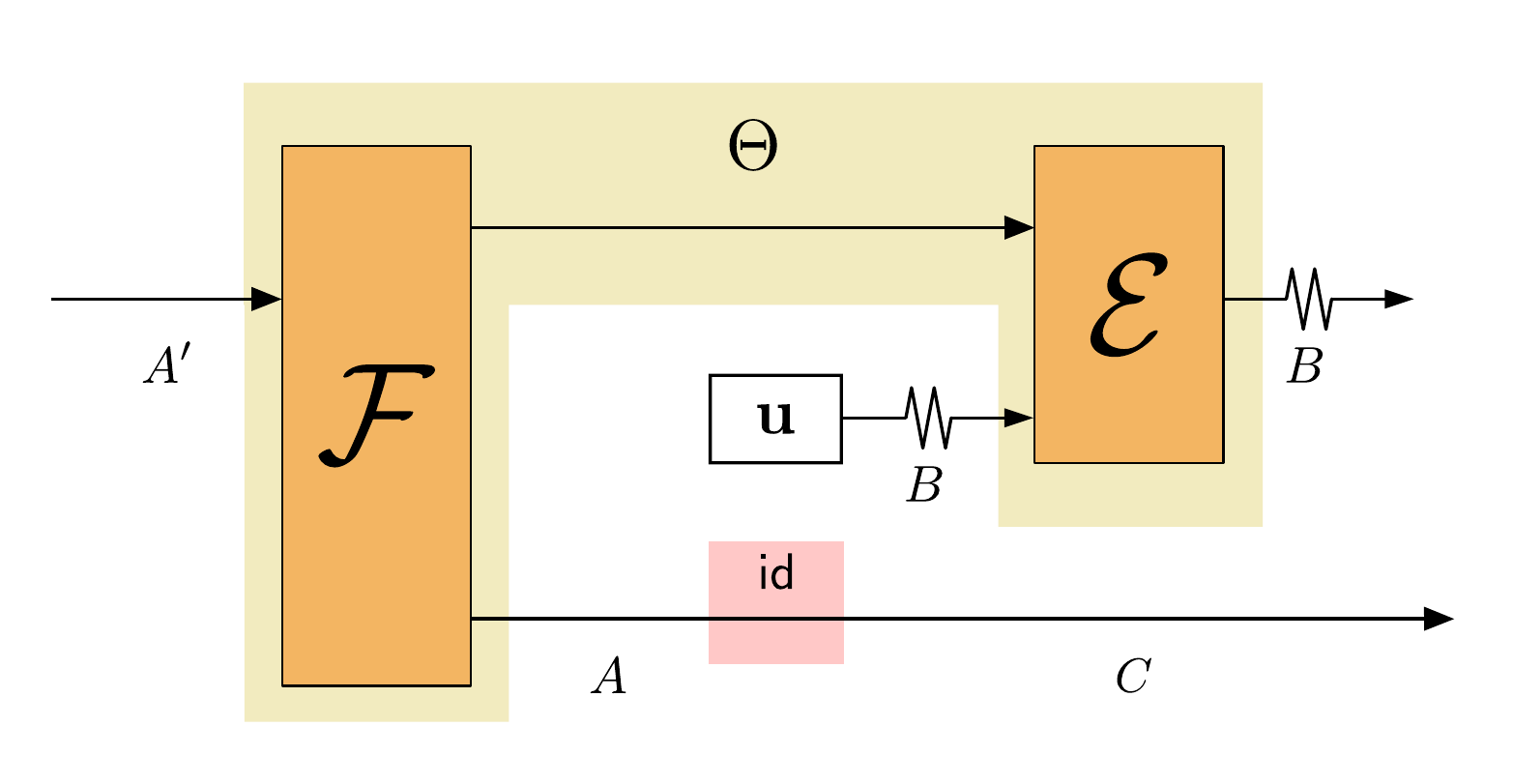}
  \caption{\small The superchannel $\Theta$ maintains the marginal uniformity of the channel $\id^{A\to C}\otimes \u^B$.}
  \label{fig:single-channel-check}
\end{figure}

At first glance, the definition of channel majorization involving the unbounded dimensions of $C$ might appear complex, particularly because it requires verifying that the channel $\mathcal{M}$ remains marginally uniform across all marginally uniform choices of $\mathcal{N}$. However, building on Choi's work, which characterizes completely positive maps by the preservation of positivity for a specific operator --- the maximally entangled state \cite{Choi1975} --- we introduce a streamlined approach. In the appendix, we demonstrate that a superchannel $\Theta$ preserves complete uniformity if and only if it maintains the marginal uniformity of a particular channel:
\be
\mN^{A\to BC}=\id^{A\to C}\otimes \u^{B}\;, 
\ee
where we took $|C_0|=1$ so that $C=C_1$  is a replica of $A$. That is, $\Theta$ is completely uniformity-preserving if the channel 
\begin{equation}
\Theta \otimes \mathds 1^C \left[\id^{A\to C}\otimes \u^{B}\right]
\end{equation} 
is marginally uniform with respect to $B$ (see Fig.~\ref{fig:single-channel-check}).

\subsubsection*{Equivalence}
The axiomatic and constructive approaches to defining mixing superchannels in the quantum domain result in the same set of operations, mirroring the situation we encountered in the classical case. This convergence is captured in the following theorem, for which we provide a proof in the appendix, but has already been proved as part of theorem 3 in~\cite{Gour2019}:
\bmyt\label{th:quantum_mixed=CUP}
A superchannel is a mixing operation if and only if it is completely uniformity preserving.
\emyt

The theorem above states that every superchannel as given in~\eqref{theta} is completely uniformity preserving if and only if $\mE$ is conditionally unital. The equivalence demonstrated by this theorem provides a compelling justification for our definition of quantum channel majorization. It shows that our intuition about mixing operations, whether approached from a constructive or axiomatic perspective, converges to a single, well-defined concept. 
\subsection{The Preorder of Channel Majorization}

Having established the equivalence between our constructive and axiomatic approaches, we can now formally define quantum channel majorization:

\bmyd\label{maind}
 We say that $\mN$ \emph{majorizes} $\mM$ and write
$\mN\succ\mM$, if there exists a mixing superchannel $\Theta$ such that
$
\mM=\Theta[\mN]
$.
\emyd
We will also use the notation $\mN\sim\mM$ if both $\mN\succ\mM$ and $\mM\succ\mN$. In this case, we will say that $\mN$ and $\mM$ are equivalent.

This definition gives rise to several fundamental properties of quantum channel majorization:

Similarly to the classical domain, if the input systems are trivial, that is, if $A\cong A' \cong \mathbb C$ the channels $\mN$ and $\mM$ are naturally identified with the states $\rho^B=\mN(1)$ and $\sigma^B= \mM(1)$, and in this case, channel majorization reduces to state majorization between $\rho$ and $\sigma$.  The set of pure states, $\pure(B)$, forms the equivalence class of maximal elements under state majorization. Consequently, pure states must also be equivalent under channel majorization.

Another basic property of channel majorization
 is that 
 \begin{equation}\label{preprocessing}
     \mN \succ \mN \circ \mF
 \end{equation}
 for any two channels $\mN \in \cptp(A\to B)$ and $ \mF \in \cptp(A'\to A)$.
This relation captures the intuition that pre-processing can only disrupt the control over the input. This implies, for instance, that a state $\rho^{B}$ is equivalent to any replacement channel of the form $\mN(\sigma^{A}) = \tr[\sigma^{A}] \rho^{B}$, as they are interchangeable via pre-processing. Thus, channel majorization between replacement channels reduces to state majorization between the channels' respective outputs.
	 
Channel majorization respects the topological, convex, and tensor structures of quantum mechanics: The subset of channels in $\cptp(A' \to B)$ that majorize $\mN$ is topologically closed and the subset of those majorized by $\mN$ is both topologically closed and convex. Moreover, if $\mN_1 \succ \mM_1$ and $\mN_2 \succ \mM_2$ then $\mN_1 \otimes \mN_2 \succ \mM_1 \otimes \mM_2$. 
 
Some preorders have maximal and minimal elements. Clearly, the uniform channel is minimal with respect to channel majorization. In the appendix, we show that the identity channel is maximal. Formally, for all systems $A$ and for all $\mN\in\cptp(A\to B)$ we have 
\be
    \id^{B}\succ\mN^{A\to B} \succ \mR^{B}\;.
\ee
From Eq.~(\ref{preprocessing}), we conclude that any channel $\mN^{A\to B}$ that has a right inverse---a channel $\mF^{B\to A}$ such that $\mN \circ \mF = \id^B$---is also maximal.
This includes unitary channels in the case $A\cong B$ and left inverses of isometries in the case $|A|>|B|$.
For the case  $|B|> |A|$, we show in the appendix that the maximal elements in $\cptp(A\to B)$ are the isometry channels.

The last property we consider here involves a combination of maximal and minimal elements. Specifically, let $B'$ be a replica of $A$ and suppose $|B|\geq |A|$. Let $\mV\in\cptp(A\to B)$ be an isometry channel and $\psi \in \pure(BB')$ the bipartite channel $\mV^{A \to B} \otimes \u^{B'}$ combines the maximal element $\mV\in \cptp(A\to B)$ and the minimal element $\u^{B'}$ of channels with output system $B'$.

\bmyt\label{paat}
Under the assumptions above, 
\be
\mV^{A\to B} \otimes \u^{ B'}\sim \psi^{BB'}\;.
\ee
\emyt

More generally, suppose $BC = B'C'$ are two bipartite representations of the output space. Then, we have 
$$\mV^{A\to B} \otimes \u^{ C}\succ \mN^{A'\to B'} \otimes \u^{ C'}\;,$$
for any channel $\mN\in \cptp{(A' \to B')}$ whenever $\mV$ is an isometry channel and $|AC'| \ge |A'C|$.

\subsection{Relationship to Conditional Majorization}
Conditional majorization~\cite{GGH+2018,GWBG2022} (see also chapter 7 of~\cite{Gour2024}) is a preorder defined on the set of bipartite density matrices. Specifically, consider two systems $A$ and $A'$ held by Alice and one system $B$ held by Bob. We say that a bipartite density matrix $\rho^{A'B}$ conditionally majorizes another bipartite state $\sigma^{AB}$ relative to Bob, and write $\rho^{A'B}\succ_B\sigma^{AB}$, if there exists a conditionally mixing channel $\mN\in\cptp(A'B\to AB)$ such that
\be
\sigma^{AB}=\mN^{A'B\to AB}\left(\rho^{A'B}\right)\;.
\ee
The channel $\mN^{A'B\to AB}$ is a conditionally mixing channel if it satisfies two conditions: 
\ben
\item It is $B\not\to A$ signaling, i.e., it can be expressed as
\be\label{semi}
\mN^{A'B\to AB}=\mE^{RB\to B}\circ\mV^{A'\to RA}
\ee
for some isometry $\mV$ and a channel $\mE$. The reference system $R$ can always be taken to have dimension 
\be
|R|=\rank\left(J^{A'A}_\mN\right)\leq |A'A|
\ee
where $J^{A'A}_\mN$ is the marginal of the Choi matrix $J^{A'BA \tB}_\mN$ of $\mN^{A'B\to A \tB}$ ($\tB$ is a replica of $B$).
\item It is conditionally unital; i.e., for all states $\rho^{B}$
\be\label{cu}
\mN^{A'B\to AB}\left(\rho^{A'}\otimes\u^B\right)=\sigma^{A}\otimes\u^{B}
\ee
where $\sigma^{A}$ is some density matrix. 
\een
Note that by tracing out system $B$, we can express $\sigma^{A}$ as
\be\label{cunital00}
\sigma^{A}= \mN^{A'B\to A}\left(\rho^{A'}\otimes\u^B\right)\;,
\ee
where $\mN^{A'B \to A} \eqdef \tr_{B} \circ \mN^{A'B \to AB}$.

In order to make the connection between conditional majorization and channel majorization we will use the following characterization of conditional mixing channels (this characterization was not discovered before, and it is proved in the appendix). 

\bmyt\label{th:condmix}
The channel $\mN^{A'B\to AB}$ is a conditionally mixing operation if and only if it has the form~\eqref{semi}, where $\mE^{RB\to B}$ is a conditionally unital channel.
\emyt

\begin{remark}
It is important to observe that the theorem specifies that a $B\not\to A$ signaling channel $\mN$, as described in~\eqref{semi}, is conditionally unital if and only if the channel $\mE$ is conditional unital. This insight offers a subtle simplification of the condition presented in equation~\eqref{cu}.
\end{remark}

With this theorem, it becomes evident that there is a close relationship between channel majorization and conditional majorization. Specifically, for every superchannel $\Theta$, defined as in equation~\eqref{theta}, we can construct the channel
\be
\Delta_\Theta^{A'B\to A B}\eqdef\mE^{RB\to B}\circ\mV^{A'\to RA}\;.
\ee
Note that the channel $\Delta_\Theta$ does not allow signaling from $B$ to $A$. Consequently, the mapping
\be
\Theta\mapsto\mathbf{f}(\Theta)\eqdef\Delta_\Theta\;,
\ee
establishes a bijection between the set of all superchannels from $\mathrm{CPTP}(A \to B)$ to $\mathrm{CPTP}(A' \to B)$, and the set of all $B\not\to A$-signaling channels in $\mathrm{CPTP}(A'B \to A B)$. Theorem~\ref{th:condmix} tells us that $\mathbf{f}$ restricts to a bijection between the set of mixing superchannels and the set of conditional mixing bipartite channels in $\mathrm{CPTP}(A'B \to AB)$.

The latter bijection indicates a potential close relationship between channel majorization and conditional majorization. 
Unfortunately, only rarely does this bijection enable automatic translation between conditional and channel majorization.
This is because, generally, channels and bipartite states have no corresponding natural identification to interplay with $\mathbf{f}$.

However, in the classical scenario, the connection is more apparent, as illustrated by the following theorem.
																															   
\bmyt\label{th:conditional-channel}
Let $\mN^{X\to Y}$ and $\mM^{X'\to Y}$ be two classical channels. Then, $\mN\succ \mM$ if and only if
\be
\sum_{x\in[m]}\!\!p_x\e_x\otimes\mN(\e_x) \!\succ_Y\!\!\!\!\!\sum_{w\in[m']}\!\!\!\!q_w \e_w\otimes\mM(\e_w)\!\!
\ee
for some $\mathbf{p}\in \prob(m)$ and $\mathbf{q}\in \prob(m')$.

\emyt

\begin{figure}
    \centering
    \includegraphics[width = \columnwidth]{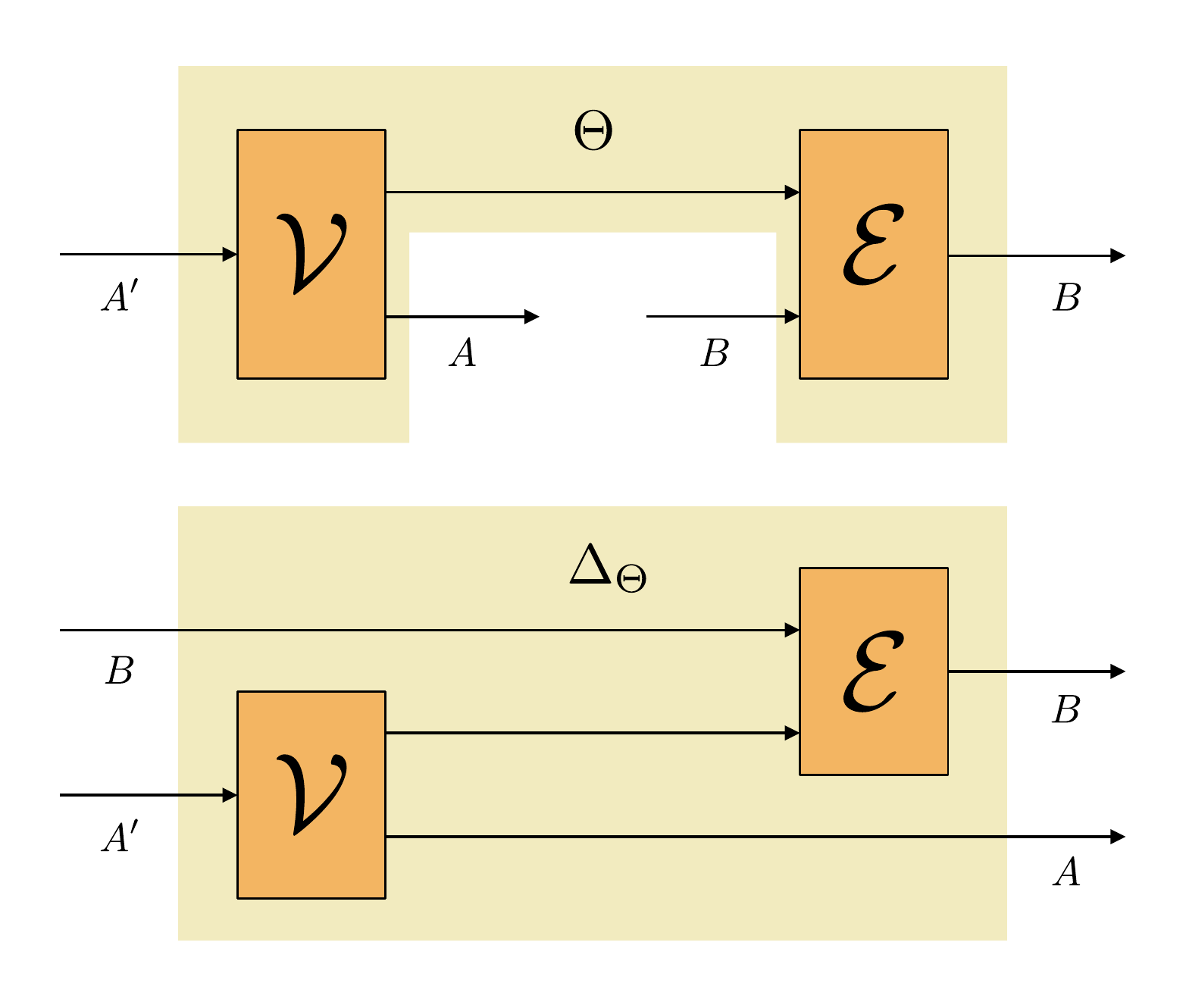}
    \caption{A bijection between mixing superchannels and conditionally mixing channels.}
    \label{fig:superchannel-channel-bijection}
\end{figure}

\subsection{The Case of Different Output Dimensions} \label{subsection:different-output-dimension}

Majorization typically involves comparing two probability vectors of equal dimensions, but this can be extended to vectors of differing dimensions by padding the shorter vector with zeros. 
We apply a similar principle of embedding one system in the other by isometries to compare channels with non-matching output dimensions as follows:

Let $\mN\in\cptp(A\to B)$ and $\mM\in\cptp(A'\to B')$  be any two channels, where we no longer assume that $B \cong B'$. We say that $\mN^{A\to B}$ majorizes $\mM^{A'\to B'}$ and write
$\mN^{A\to B}\succ\mM^{A'\to B'}$, if one of the following occurs:
\ben 
\item $|B|>|B'|$ and there exists 
an isometry channel $\mU \in \cptp({B'\to B})$  such that 
\be
\mN^{A\to B}\succ\mU^{B'\to B}\circ\mM^{A'\to B'}
\ee
where $\succ$ in the equation above stands for channel majorization as introduced in Definition~\ref{maind} for two channels of the same output dimensions.
\item $|B|\le|B'|$ and there exists an isometry channel $\mV \in \cptp({B\to B'})$ such that 
\be
\mV^{B\to B'}\circ\mN^{A\to B}\succ\mM^{A'\to B'}\;.
\ee
\een
																																						   
This generalization of channel majorization allows us to define channel entropy as a function across channels of all output dimensions.

\section{Channel Entropy}
\label{sec:entropy-quantum-channel}

In the work by \cite{GT2021}, entropy for classical static systems is characterized as a function over probability vectors that exhibits Schur-concavity and additivity. \footnote{This axiomatic approach to entropy heralds back to the axiomatization of classical thermodynamics by Lieb and Yngvason, which constructs classical thermodynamic entropy to be an additive monotone of the preorder of adiabatic accessibility. \cite{LY1999}} Following from the findings of \cite{MPS+2021}, any function that meets these criteria necessarily represents a convex combination of R\'enyi entropies. In this section, we aim to generalize this well-founded definition of entropy to the domain of channels in the most logical and straightforward manner. This extension will allow us to explore entropy in the broader context of channels, aligning the conceptual underpinnings of static entropy of states with dynamic quantum processes.

\subsection{Definition and General Properties}

Let
\be\label{tff}
\H:\bigcup_{A,B}\cptp(A\to B)\to\mathbb{R}\;,
\ee
be a mapping of quantum channels between systems of finite dimensions to the real line. $\H$ assigns to each quantum channel $\mN^{A\to B}$ a real number, denoted by $\H(B|A)_\mN$. Note that we are using similar notations as used for conditional entropy since the entropy of a quantum channel signifies the uncertainty associated with system $B$ given complete control over the input of the channel.

Our objective is to identify when $\H$ constitutes an entropy. For systems where $|A|=1$, we denote $\H(B|A)_\mN$ as $\H(B)_\mN\eqdef\H(\rho^{B})$, where $\rho^B := \mN(1)$ aligning with the notation of quantum-state entropy. This notation proves useful when examining composite systems with multiple subsystems. Since we define entropy functions as non-constant zero functions, we assume (implicitly throughout this paper, and in the definition below) the existence of non-trivial systems $A$ and $B$ and a channel $\mN\in\cptp(A\to B)$ such that $\H(B|A)_\mN\neq 0$.

\bmyd
    The function $\H$ as given in~\eqref{tff} is called a \emph{channel entropy} if it satisfies the following properties:
    \ben
    \item Monotonicity: If $\mN^{A\to B}\succ\mM^{A'\to B'}$ then $\H(B|A)_{\mN}\leq\H(B'|A')_{\mM}$,
    \item Additivity: $\H(BB'|AA')_{\mN\otimes\mM}=\H(B|A)_{\mN}+\H(B'|A')_\mM$.
    \een
\emyd
\begin{remark}We say that a quantum channel entropy is \emph{normalized} if the uniform qubit state has unit entropy.
The term `normalized' is justified in the appendix, where we prove in a similar fashion to~\cite{GWBG2022} that any channel entropy must map the uniform qubit state to a strictly positive number.
Henceforth, we shall also implicitly assume that channel entropies are normalized.
\end{remark}

There are several properties of entropy that follow directly from the definition above. First, observe that in the case that the input system $A$ is trivial (when $|A|=1$), a channel entropy function reduces to a state entropy function.
Specifically, irrespective of the normalization axiom, every pure state necessarily has zero entropy, while every mixed state has positive entropy.
Secondly,  the entropy of a replacement channel equals the entropy of the state that it outputs, as intuitively expected.

Next, the entropy of quantum channels is invariant under the action of pre-processing unitaries and post-processing isometries.
That is, for a quantum channel~$\mN^{A\to B}$, unitary channel $\mathcal{U}^{A'\to A}$, and isometry channel $\mathcal{V}^{B\to B'}$,
\be\label{eq:secoiu}
    \H(B|A)_{\mN} = \H(B'|A')_{\mV\circ\mN\circ\mU}\;.
\ee
From the normalization and additivity properties, it follows that for any two systems $A$ and $B$, $\H(\mR^{A\to B}) = \log |B|$.
Therefore, from Theorem~\ref{paat}, we deduce 
\be\label{addi}
0= \log|A| + \H(B|A)_{\mV}\;,
\ee
for any isometry channel $\mV^{A\to B}$.
That is, isometries with an input dimension greater than one have negative entropy. We summarize this key observation in the following theorem.
\bmyt \label{inevitability}
    Let $\H$ be a channel entropy and $\mV\in\cptp(A\to B)$ be an isometry channel. Then,
         \be
         \H(B|A)_{\mV} = - \log |A|\;.
         \ee
\emyt

\subsection{Examples of Channel Entropies}

One set of examples of channel entropies are those derived from channel divergences.  Specifically, if $\D$ be a relative entropy of channels.~\cite{GW2021}, then the function
\be\label{48}
\H(B|A)_{\mN}\eqdef\log|B|-\D\left(\mN\big\|\mR\right)
\ee
is an entropy. The additivity follows from the additivity of $\D$ and the monotonicity under mixing operations from the data processing inequality of $\D$. In fact, as discussed also in~\cite{GW2021}, this example of channel entropy behaves monotonically under the larger set of uniformity-preserving superchannels.

There is another way to construct channel entropies from conditional entropies of bipartite quantum states.
Specifically, a conditional entropy $\h:\rho^{RB}\mapsto\h(B|R)_\rho$ induces a measure of quantum channels $\H$ by:
\be\label{inf2}
\H(B|A)_{\mN}\eqdef\inf\h\left(B|R\right)_{\mN^{A\to B}\left(\rho^{RA}\right)}\;,
\ee
where the infimum is over all systems $R$ and all density matrices $\rho^{RA}$. We first argue that we can take the optimization above to run over all pure states $\rho^{RA}$ with $R\cong A$. To see why, first recall that if $\psi^{R'RA}$ is a purification of $\rho^{RA}$ then
\be
\h\left(A|R\right)_{\mN^{A\to B}\left(\rho^{RA}\right)}\geq \H\left(B|RR'\right)_{\mN^{A\to B}\left(\psi^{R'RA}\right)}\;,
\ee
since the tracing of $R'$ in this context is a conditionally mixing operation (and conditional entropies of bipartite states behave monotonically under such operations~\cite{GWBG2022,Gour2024}). Thus, we can restrict the infimum in~\eqref{inf2} to pure states. Moreover, since conditional entropy is invariant under isometries, we get from the Schmidt decomposition of $\psi^{R'RA}$  that we can replace system $R'R$ with a system isomorphic to $A$. We therefore conclude that
\be
\H(B|A)_{\mN}\eqdef\min_{\psi\in\pure(RA)}\h\left(B|R\right)_{\mN^{A\to B}\left(\psi^{RA}\right)}\;,
\ee
where $R$ is a replica of $A$. In the appendix, we show that $\h(B|A)_{\mN}$ as defined above satisfies the monotonicity axiom of entropies, and we give now two examples in which it satisfies the additivity axiom as well so that for these cases it is indeed a channel entropy.

When $\D$ is taken to be the Umegaki relative entropy, and $\H$ is taken to be the von-Neumann conditional entropy we get the same channel entropy
~\eqref{48} and in~\eqref{inf2}. While this equality does not hold in general, it does happen also for the min-channel entropy. That is, substituting $\D=D_{\max}$ (the max-relative entropy) in~\eqref{48} yields the same channel entropy as we get by taking $\H=H_{\min}$ to be the conditional min-entropy.

Specifically, 
the min-entropy of a quantum channel can be defined similarly to~\eqref{48} as:
\ba \label{min-entropy-cp}
H_{\rm min}\left(B|A\right)_{\mN}&=\log|B|-D_{\max}\left(\mN\big\|\mR\right)\\
&=\log|B|-\min\big\{\log t\;:\; t\mR\geq \mN\big\}\;,
\ea
 where the relation $\ge$ between superoperators is the partial order induced by the convex cone of completely positive maps ($t\mR\geq \mN$ if and only if $t\mR- \mN$ is completely positive).
Or similar to~\eqref{inf2} as:
\ba \label{min-entropy-pure}
{}&H_{\min}(B|A)_{\mN}\eqdef\min_{\psi\in\pure(RA)}H_{\min}\left(B|R\right)_{\mN^{A\to B}\left(\psi^{RA}\right)}\\
&=-\max_{\psi\in\pure(RA)}D_{\max}\left(\mN^{A\to B}(\psi^{RA})\big\|\psi^R\otimes I^B\right)\;.
\ea
It is not too hard to check that the two definitions above coincide (see appendix for more details). The following theorem provides justification for the terminology of the min-entropy of a channel.

\bmyt\label{hmin-justification}
Let $\H$ be a channel entropy and $\mN\in\cptp(A\to B)$, then
\be
\H(B|A)_{\mN}\geq H_{\min}(B|A)_{\mN}.
\ee
\emyt
Observe that we have already established equality between all channel entropies whenever $\mN$ is either a uniform or an isometry channel.

\subsection{Entropy of Classical Channels}

The entropy of classical channels can be viewed as extensions of entropy functions that are defined on probability vectors. If $\h$ is an entropy of a classical probability vector, we can define its maximal and minimal extensions to the classical channel domain using the approach introduced in~\cite{GT2020}. Specifically, we define the minimal and maximal extensions of $\h$ to the channel $\mN\in\cptp(X\to Y)$ as follows: 
\ba
    \underline{\h}(Y|X)_\mN&\eqdef\sup_{\q}\Big\{\h(\q)\;:\;\q\succ\mN\Big\}\label{eq:min-extension}\;,
    \\
\overline{\h}(Y|X)_\mN&\eqdef\inf_{\q}\Big\{\h(\q)\;:\;\mN\succ\q\Big\}\;, 
\ea
where the infimum/supremum is over all the probability vectors of arbitrary finite length. Note that the minimal extension of an entropy is super-additive, while the maximal extension is sub-additive. That is
\begin{align*}
    \underline{\h}(YY'|XX')_{\mN \otimes \mM} &\ge \underline{\h}(Y|X)_\mN + \underline{\h}(Y'|X')_\mN \quad \text{and}\\
    \overline{\h}(YY'|XX')_{\mN \otimes \mM} &\le
    \overline{\h}(Y|X)_\mN + \overline{\h}(Y'|X')_\mN\;,
\end{align*}
Every entropy function on classical channels $H'$ that reduces to $\h$ on probability vectors must satisfy
\be\label{58}
\underline{\h}(Y|X)_\mN\leq H'(Y|X)_\mN\leq 
\overline{\h}(Y|X)_\mN\;.
\ee
for every classical channel $\mN^{X\to Y}$. In particular, note that both bounds are non-negative and hence so must be $H'$.

In the next theorem, we provide a formula for the maximal extension for the case when $\h$ is quasiconcave (e.g.~R\'{e}nyi entropies). As before, we
denote by $\p_1,\ldots\p_m\in\prob(n)$ the output vectors of $\mN$. 
\bmyt\label{th:maxext}
Using the notations above,
\be
\overline{\h}(Y|X)_\mN=\min_{x\in[m]}\h(\p_x)
\ee										   
for any quasiconcave classical entropy $\h$.
\emyt
That is, the maximal extension of a quasiconcave entropy function equals its minimum entropy output \cite{KR2001}. In the classical case, it is known that the minimum entropy output is additive. Thus, $\overline{\h}(Y|X)_\mN$ is an entropy. However, in general, minimal and maximal extensions of a function from a smaller domain to a larger one may not be additive even if the function is additive on the restricted domain. Indeed, as we discuss now the minimal extension $\underline{\h}(Y|X)_\mN$ is \emph{not} additive.

In Example~1 of Sec.~\ref{sec:g} we discussed the optimal (i.e., minimal) vector $\r \in \prob(n)$ satisfying $\r \succ \mN$. Almost immediately from the definition we have $\underline{\h}(Y|X)_\mN=\h(\r)$. Although the expression for the components of the optimal vector $\r$ is somewhat cumbersome (see appendix), it is relatively simple to show that it is not additive, and therefore, neither is $\underline{\h}(Y|X)_\mN$. Thus, we conclude that the minimal extension is not an entropy.

While $\underline{\h}(Y|X)_\mN$ is not additive under tensor products, it is super-additive, and therefore can be regularized. That is, its regularization is defined as
\be\label{61}
\underline{\h}^\reg(Y|X)_\mN\eqdef \lim_{n\to\infty}\frac1n\underline{\h}(Y^n|X^n)_{\mN^{\otimes n}}\;,
\ee
where Fekete's Lemma ensures that the super-additivity of $\underline{\h}$ implies the existence of the above limit. Observe that $\underline{\h}^\reg(Y|X)_\mN$ is at least weakly additive in the sense that for all $k\in\mbb{N}$ we have
\be\label{62}
\underline{\h}^\reg\left(Y^k\big|X^k\right)_{\mN^{\otimes k}}=k\underline{\h}^\reg(Y|X)_\mN\;.
\ee
It is left as an open problem to verify whether this regularized version of $\underline{\h}$ is fully additive so that it is an entropy function. Remarkably, the minimal and maximal extensions of the Shannon entropy, $H(\p)\eqdef-\sum_{x\in[n]} p_x\log p_x $, are equal and, therefore, must be fully additive. (cf. Theorem $3$ of \cite{BGG2022})
\bmyt\label{19t}
Let $\h$ be a channel entropy that reduces to the Shannon entropy of probability vectors. Then for every classical channel $\mN$,
\ba
	\h(Y|X)_\mN &= \underline{H}^\reg(Y|X)_\mN\\&=\overline{H}(Y|X)_\mN\\&=\min_{x \in [m]} H(\p_x).
\ea
\emyt

This theorem serves as a uniqueness result, indicating that there is only one classical channel entropy that reduces to the Shannon entropy on probability vectors.

\subsection{Quantum Additivity Implies Negativity}

In Theorem~\ref{inevitability}, we have learned that every entropy of quantum channels must assume strictly negative values. This is not the case for classical channels since in~\eqref{58} we saw that every entropy function on classical channels that reduces to $\h$ on probability vectors must be no smaller than its minimal extension $\underline{\h}$ as defined in~\eqref{eq:min-extension}. Thus, since $\underline{\h}$ is non-negative, every entropy of classical channels is always non-negative.

At first glance, it may seem that the same argument should work also for the quantum domain. That is, let $\h$ be an entropy on classical states (probability vectors) and let $\underline{\h}$ be its minimal extension on a quantum channel $\mN\in\cptp(A\to B)$ given as
\be
\underline{\h}(B|A)_\mN\eqdef\sup_{\q}\Big\{\h(\q)\;:\;\q\succ\mN\Big\}\;.
\ee
where the infimum/supremum is over all the probability
vectors of arbitrary finite length. The problem with the function above is that it is not well defined for many quantum channels in $\cptp(A\to B)$. That is, for many quantum channels there is no classical probability vector $\q$ that satisfies $\q\succ\mN$. Thus, for such channel $\mN$, we cannot argue that its entropy is non-negative.

Another attempt to construct a non-negative entropy of a quantum channel may involve the maximal extension defined as follows. As before, let $\h$ be an entropy on classical states and let $\overline{\h}$ be its maximal extension on a quantum channel $\mN\in\cptp(A\to B)$ given as
\be
\overline{\h}(B|A)_\mN\eqdef\inf_{\q}\Big\{\h(\q)\;:\;\mN\succ\q\Big\}\;.
\ee
Unlike the minimal extension, this maximal extension is well defined for every $\mN\in\cptp(A\to B)$ since every quantum channel majorizes the uniform state which can be viewed as a classical state (probability vector).
																															  
While the function $\overline{\h}(B|A)_\mN$ behaves monotonically under channel majorization, it is not additive and therefore not an entropy. Thus, if we remove the requirement of additivity from the definition of entropy, we would get functions that can be non-negative on all quantum channels. 

Since the function $\overline{\h}(B|A)_\mN$ is sub-additive, its regularization exists and is given by
\be
\overline{\h}^\reg(B|A)_\mN\eqdef \lim_{n\to\infty}\frac1n\overline{\h}(B^n|A^n)_{\mN^{\otimes n}}\;.
\ee
Similar to~\eqref{62}, the function above is weekly additive in the sense that for all $k\in\mbb{N}$ we have
\be
\overline{\h}^\reg\left(B^k\big|A^k\right)_{\mN^{\otimes k}}=k\overline{\h}^\reg(B|A)_\mN\;.
\ee
Thus, the function $\overline{\h}^\reg$ has the following three properties:
\ben
\item It is non-negative.
\item It satisfies the monotonicity axiom of an entropy.
\item It satisfies weak additivity.
\een
Moreover, combining this with Theorem~\ref{inevitability} we get that $\overline{\h}^\reg$, in general, \emph{cannot} be (fully) additive.
Thus, it is the strong (i.e., non-weak) additivity axiom of entropy of quantum channels that gives rise to negative values. In fact, recall that we used strong additivity to get~\eqref{addi}. 

\section*{Conclusions}

In this study, we meticulously crafted a framework for both classical and quantum channel majorization, approached through constructive, axiomatic, and operational methodologies. Despite their methodological distinctions, each approach culminates in a consistent and robust ordering of channels. Significantly, our findings demonstrate that any entropy function applicable to quantum channels must adopt negative values. This discovery not only challenges conventional understandings of entropy but also necessitates a reevaluation of its foundational concepts within quantum mechanics.

Looking forward, several avenues exist for expanding this research. Initially, we defined games of chance exclusively within classical channels. Extending this to quantum channels remains a compelling prospect. While preliminary efforts have been made~\cite{BGWG2021}, identifying an operational framework that incorporates quantum games of chance and aligns with the ordering derived from quantum channel majorization presents an intriguing challenge.

Additionally, previous work provided operational interpretations for the von-Neumann entropy extension to quantum channels, specifically as the optimal rate in a quantum channel merging protocol. Further exploration into operational interpretations for both von-Neumann entropy and other Rényi entropies could yield new insights.

In the domain of classical channels, an open question persists regarding which state entropies, $\h$, see their regularized minimal extensions align with their maximal extensions. In Theorem~\ref{19t} we established that for the Shannon entropy, this alignment offers a uniqueness result, demonstrating a unique extension of Shannon entropy to classical channels. Whether similar conclusions hold for R\'enyi entropies remains to be seen.

Ultimately, this research clarifies the framework of quantum channel majorization and opens avenues for future explorations into the operational manipulation of quantum channels. Inspired by the success of the quantum state merging protocol, where negative entropy plays a pivotal role, this concept, particularly within this framework, promises to drive significant advancements in both the theoretical and practical realms of quantum information science.

\begin{acknowledgments}
GG wishes to express gratitude to Mark Wilde for engaging in valuable discussions related to the topics covered in this paper. Additionally, GG thanks Ludovico Lami for suggesting the key terms ``constructive," ``axiomatic," and ``operational," which have significantly shaped this work. GG, DK, and TN acknowledge support from NSERC.
\end{acknowledgments}
\bibliography{ne10-paper}

\onecolumngrid
\newpage
\appendix
\section*{Appendix}
\subsection{The standard form of a superchannel}

\begin{lemma}\label{lm:superchannelstandardform}
    Every classical superchannel $\Theta$ can be written in the following \emph{standard form}
    \be\label{eq:standardform}
    \Theta\left[\mN^{X \to Y}\right]=\sum_{\substack{x\in[m]\\{w\in[m']}}}\mE^{Y\to Y'}_{xw}\circ\mN^{X\to Y}\circ\mF^{X'\to X}_{xw}
    \ee
    where for each $x\in[m]$ and $w\in[m']$, $\mE_{xw}^{Y\to Y'}$ are classical channels, and for all $w' \in [m']$
    \be\label{eq:classicalpreprocessing}
        \mF^{X'\to X}_{xw}\left(\e^{X'}_{w'}\right)\eqdef\delta_{ww'}s(x\vert w)\e_x^{X},
    \ee
    where $\{s(x\vert w)\}_{x,w}$ is some conditional probability distribution.
\end{lemma}
\begin{proof}
Any classical superchannel $\Theta$ can be realized as 
\be
\Theta\left[\mN^{X\to Y}\right]=\mE^{YZ\to Y'}\circ\mN^{X\to Y}\circ\mS^{X'\to XZ}
\ee
for some classical channels $\mS^{X' \to XZ}$ and $\mE^{YZ \to Y'}$. Letting $k \eqdef |Z|$, we have that $\mE_z^{Y \to Y'}(\e_y^{Y}) \eqdef \mE^{YZ \to Y'}(\e_y^{Y} \otimes \e_z^{Z})$ and $\mS_z^{X' \to X}(\e_w^{X'}) \eqdef \sum_{x \in [m]}s(xz|w)\e_x^{X}$ for each $z \in [k]$, where the conditional probability distribution $\{s(xz|w)\}_{x,z,w}$ are given by
\be
S^{X' \to XZ}(\e_w^{X'}) = \sum_{x \in [m]}\sum_{z \in [k]}s(xz|w)\e_x^{X} \otimes \e_z^{Z}\;,
\ee
we have
\be\label{000}
\Theta\left[\mN^{X\to Y}\right]=\sum_{z\in[k]}\mE_{z}^{Y\to Y'}\circ\mN^{X\to Y}\circ\mS_z^{X'\to X}\;.
\ee
It follows that
\be
\Theta\left[\mN^{X\to Y}\right]\left(\e_w^{X'}\right)=\sum_{x\in[m]}\sum_{z\in[k]}s(xz|w)\mE_{z}^{Y\to Y}\circ\mN^{X\to Y}\left(\e_x^{X}\right)\;.
\ee
Now, denote by $s(x|w)\eqdef\sum_{z\in[k]}s(xz|w)$ and define the channels 
\be
\mE^{Y\to Y'}_{xw}\eqdef\frac{1}{s(x|w)}\sum_{z\in[k]}s(xz|w)\mE_{z}^{Y\to Y'},
\ee
for the case that $s(x|w)>0$, and otherwise $\mE^{Y\to Y'}_{xw}\eqdef 0$ for $x$ and $w$ with $s(x|w)=0$.
With these notations we get
\be\label{111}
\Theta\left[\mN^{X\to Y}\right]\left(\e_w^{X'}\right)=\sum_{x\in[m]}s(x|w)\mE_{xw}^{Y\to Y'}\circ\mN^{X\to Y}\left(\e_x^{X}\right)
\ee

Observe that for each $x\in[m]$ and $w\in[m']$ we can define a non-negative map $\mF_{xw}\in\cp(X'\to X)$ via
\be\label{eq:pre-processing_standard}
\mF_{xw}^{X'\to X}\left(\e_{w'}^{X'}\right)=\delta_{w'w}s(x|w)\e_x^{X}\;.
\ee
With this notation, \eqref{111} implies that
\be
\Theta\left[\mN^{X\to Y}\right]=\sum_{x,w}\mE_{xw}^{Y\to Y'}\circ\mN^{X\to Y}\circ\mF^{X'\to X}_{xw}\;.
\ee
\end{proof}

\subsection{Example of a uniformity-preserving but not completely-uniformity preserving superchannel}

\noindent Let $|X|=|Y|=|Y'|=2, |X'|=1$, and consider classical channels $\mS_1^{X' \to X}$, $\mS_2^{X' \to X}$, $\mE_1^{Y \to Y'}$, $\mE_2^{Y \to Y'}$ with their coresponding transition matrices
\ba
    S_{1} = \frac{1}{2} \begin{bmatrix}1 \\ 0\end{bmatrix},\qquad S_{2} = \frac{1}{2} \begin{bmatrix}0 \\ 1\end{bmatrix},\qquad E_{1} = \begin{bmatrix}1 & 1 \\ 0 & 0\end{bmatrix},\qquad E_{2} = \begin{bmatrix}0 & 0 \\ 1 & 1\end{bmatrix}.
\ea
The classical superchannel $\Theta$ defined by
\be
    \Theta[\mathcal{\mN}^{X \to Y}]\eqdef\sum_{z\in[2]} \mE_{z}^{Y \to Y'} \circ \mN^{X \to Y} \circ \mS_{z}^{X' \to X}
\ee
is uniformity preserving since $\Theta[\mR^{X \to Y}] = \mR^{X' \to Y'}$.
However, observe that $\Theta$ is not \textit{completely} uniformity preserving. For $|Z_0|=1$, $|Z_1|=2$, and the marginally uniform channel $\mN^{XZ_0 \to YZ_1}:= \u^Y \otimes I^{XZ_0 \to Z_1}$, we have
\be
    \Theta \otimes \mathds{1}^Z[\mN^{XZ_0 \to YZ_1}](\e_1^{X'Z_0}) = \frac{1}{2}\begin{bmatrix}1 \\ 0 \\ 0 \\ 1\end{bmatrix} \neq \u^{Y'} \otimes \p^{Z_1}
\ee
for any $\p \in \prob(2)$, which shows that $\Theta \otimes \mathds{1}^Z[\mN^{XZ_0 \to YZ_1}]$ is not a marginally uniform channel.

\subsection{Proof of Theorem 4}
\begin{thm}{4}
Given two classical channels $\mN$ and $\mM$, the following are equivalent:
\ben
    \item $\mN$ majorizes $\mM$ constructively.
    \item $\mN$ majorizes $\mM$ axiomatically.
    \item $\mN$ majorizes $\mM$ operationally.
\een
\end{thm}

To prove this theorem, we divided the proof into two lemmas. First, we show that the set of completely uniformity-preserving superchannels and the set of random permutation superchannels are identical. This implies (and is stronger than) $1 \iff 2$. Secondly, we show that constructive majorization is equivalent to operational majorization defined via games of chance, which proves $1 \iff 3$.

\begin{lemma}\label{lm:cup-rand-perm}
    The set of completely uniformity-preserving superchannels and the set of random permutation superchannels coincide.
\end{lemma}
\begin{proof}
    To begin with, suppose $\Theta$ is completely uniformity preserving. Then $\Theta$ preserves every marginally uniform channel $\mathcal{N}^{XZ_0 \to YZ_1}$. In particular, consider a classical channel $\mN^{XZ_0 \to YZ_1}$ where $Z_0 \to Z_1$ is a static system with trivial input system, i.e.\ $|Z_0|=1$ and output system $Y\cong X$.
    Define the channel $\mN$ as follows (see Fig.~\ref{cupid}):
    \begin{equation}
        \label{333}
        \mN^{X \to YZ_1}(\e_x^{X})\eqdef \u^{Y}\otimes\e_x^{Z_1}\quad\quad\forall\;x\in[m]\;.
    \end{equation}
    \begin{figure}[h]
        \centering    
        \includegraphics[width=0.55\textwidth]{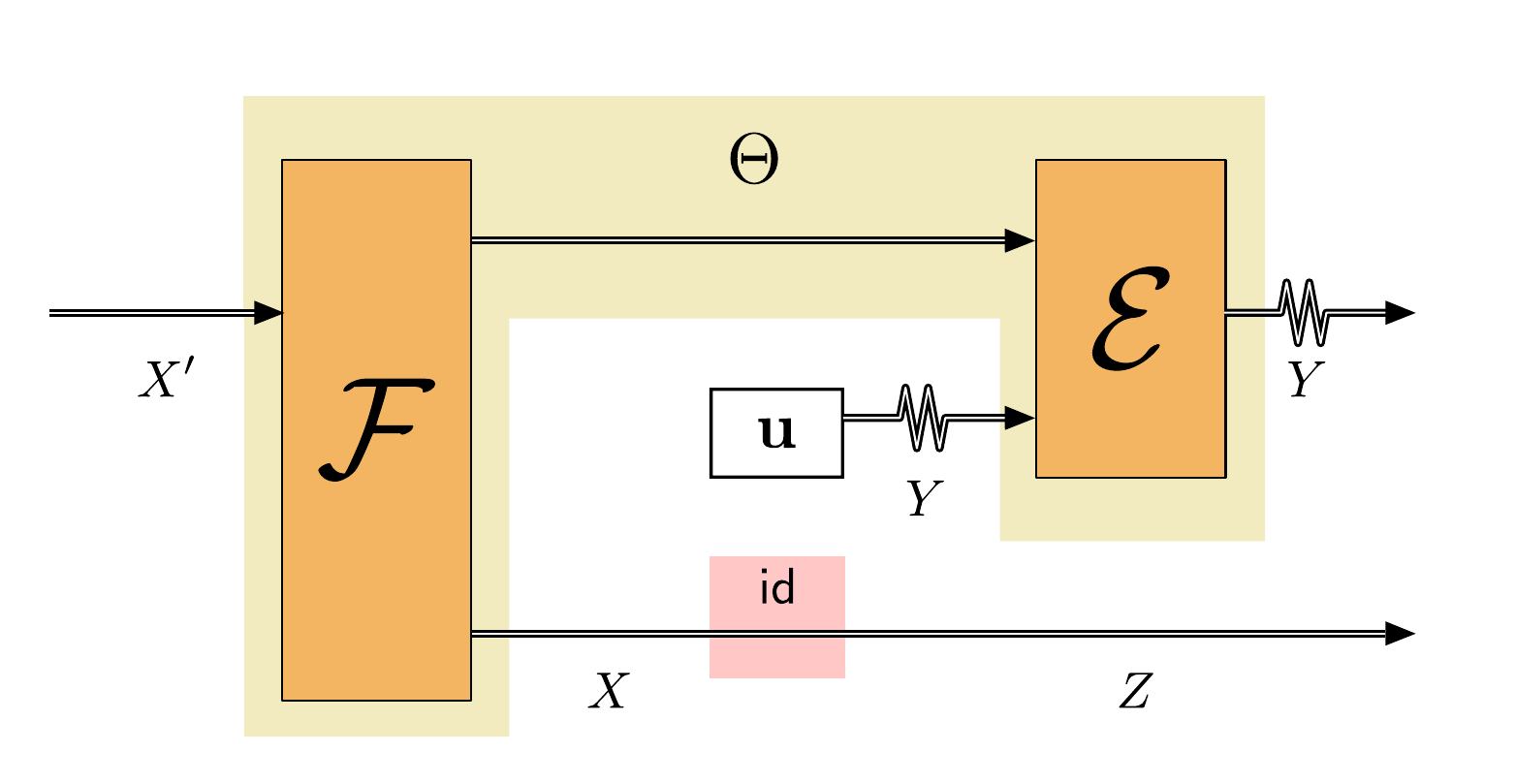}
        \caption{\small A completely uniformity-preserving superchannel preserves a channel $\u^{Y}\otimes \mathrm{id}^{X\to Z}$.}
        \label{cupid}
    \end{figure} 

    Since $\mN^{X \to YZ_1}$ is marginally uniform and $\Theta$ is completely uniformity preserving, the channel $\mM^{X' \to YZ_1}\eqdef\Theta \otimes \mathds{1}^Z \left[\mN^{X \to YZ_1}\right]$ is marginally uniform. Consider any input $\e^{X'}_y$ where $y\in[m']$,
    \ba\label{eq:cup-input}
    \mM^{X' \to YZ_1}\left(\e_{y}^{X'}\right)&=\sum_{x,w}\mE_{xw}^{Y\to Y}\circ\mN^{X \to YZ_1}\circ\mF^{X'\to X}_{xw}\left(\e_{y}^{X'}\right)\\
    &=\sum_{x\in[m]}s({x|y})\mE_{xy}^{Y\to Y}\circ\mN^{X \to YZ_1}\left(\e_x^{X}\right)\\
    \eqref{333}\rightarrow &=\sum_{x\in[m]}s(x\vert y)\mE_{xy}^{Y\to Y}\left(\u^{Y}\right)\otimes\e_x^{Z_1}
    \ea
    To show that each $\mE_{xy}^{Y\to Y}$ is doubly stochastic, we will show that each vector $\q_{xy}^{Y}\eqdef\mE_{xy}^{Y\to Y}\left(\u^{Y}\right)$ equals the uniform vector $\u^{Y}$. Since $\mM^{X' \to YZ_1}$ is marginally uniform we get from the equation above that for all $y\in[m']$
    \be
    \u^{Y}\otimes\r^{Z_1}=\sum_{x\in[m]}s(x\vert y)\q_{xy}^{Y}\otimes\e_x^{Z_1}\;,
    \ee
    for some $\r^{Z_1}\in\prob(m)$. Finally, let $x'\in[m]$, and multiply from the left both sides of the equation above by $I^{Y}\otimes \left(\e_{x'}^{Z_1}\right)^T$ (i.e., taking the dot product with $\e_{x'}^{Z_1}$ on both sides) we get
    \be
    \left(\e_{x'}^{Z_1}\boldsymbol{\cdot}\r^{Z_1}\right)\u^{Y}=s({x'|y})\q_{x'y}^{Y}\;.
    \ee
    When $s_{x'\vert y}$ is not zero, each $\q_{x'y}^{Y}$ is proportional to $\u^{Y}$ but since it is a probability vector (in particular, normalized) it must be equal to $\u^{Y}$. This shows that a completely uniformity-preserving superchannel is a random permutation superchannel. 
    
    Conversely, a random permutation superchannel is also a completely uniformity-preserving superchannel. To see this, it is enough to show this for the marginally uniform channel $\mN^{X \to YZ_1}$ defined in \eqref{333}, since any marginally uniform channel is some linear combination of channels of this form. Any input $\e^{X'}_y$ into $\Theta[\mN]$ results in
    \begin{align*}
    \Theta[\mN]\left(\e_{y}^{X'}\right)&=\sum_{x,w} \mE_{xw}^{Y\to Y}\circ\mN^{X \to YZ_1}\circ\mF^{X'\to X}_{xw}\left(\e_{y}^{X'}\right)\\
    \eqref{eq:cup-input}\rightarrow&=\sum_{x\in[m]}s(x\vert y)\mE_{xy}^{Y\to Y}\left(\u^{Y}\right)\otimes\e_x^{Z_1}.
    \end{align*}
    Since $\mathcal{E}$ is unital, it preserves the uniform vector
    \begin{equation}
        \Theta[\mN]\left(\e_{y}^{X'}\right) = \u^{Y} \otimes \sum_{x\in[m]} s(x\vert y)\e_{x}^{Z_1}
    \end{equation}
    which shows that $\Theta[\mN]$ is a marginally uniform channel. Consequently, $\Theta$ is a completely uniformity-preserving superchannel. Therefore, a random permutation superchannel is a completely uniformity-preserving superchannel.
\end{proof}
Next step is to prove the equivalence between the constructive and operational approach. To assist this, we will prove a characterization lemma.
\begin{lemma}\label{lm:char-predictability-random-perm}
    Suppose $\mN$ and $\mM$ are a pair of classical channels. The following statements are equivalent.
    \begin{enumerate}
        \item There exists a random permutation superchannel $\Theta$ such that $\mM =\Theta [\mN]$
        \item There exists an $m\times m'$ stochastic matrix $S=(s_{x|w})$ such that
        \be
            \sum_{x\in[m]}s_{x|w}\p_{x}\succ \q_{w}\quad\quad\forall\;w\in[m']\;.
        \ee
        \item For all $\s\in\prob^\da(n)$,
        \begin{equation}
        \max_{x\in[m]} \s \cdot \p^\da _x \geq \max_{w\in[m']} \s\cdot \q_w ^\da 
        \end{equation}
        where $\p_x = \mN(\e ^X _x)$, $\q_w = \mM(\e ^X _x)$ and the $\p^\da _x$ denote the rearrangment of the elements of $\p_x$ in non-increasing order.
    \end{enumerate}
\end{lemma}

\begin{proof}
        $1\implies 2$\\
       Consider the action of this output channel in the standard form of superchannel, 
        \begin{align}
            \Theta[\mN](\e_w) &=\sum_{x\in[m]}\sum_{y\in[m']} \mD_{xy}\circ\mN^{X \to Y}\circ\mS_{xy}(\e_w)\\
            &=\sum_{x\in[m]}\sum_{y\in[m']} \delta_{yw} s_{x\vert y}\mD_{xy}\circ\mN^{X \to Y}(\e_x)\\
            &=\sum_{x\in[m]} s_{x\vert w}\mD_{xw}\p_x.
        \end{align}
        Now since $\p_x\succ \mD_{xw}\p_x$ for any $x,w$, multiply both sides by $s_{x\vert w}$ and summing over $x$ yield that for any $w$,
        \begin{equation}\label{eq:majorization}
            \sum_{x\in[m]} s_{x\vert w} \p_x \succ \sum_{x\in[m]} s_{x\vert w} \mD_{xw} \p_x = \q_w.
        \end{equation}
        $2\implies 3$\\
         Suppose $\p_x = \p^\da _x$ and $\q_w = \q^\da _w$. We express the condition in 2 as
        \begin{equation}
            \sum_{x\in[m]} s_{x\vert w} L\p_x \geq L \q_w\quad \forall w \in [m']
        \end{equation}
        where $L$ is the lower triangular matrix whose lower triangular has entries one. Let $\s_{w} =(s_{1\vert w}, \dots, s_{m\vert w})^T$, we have 
        \begin{equation}\label{eq:ky-fann-lower-triang-majorization}
            LN\s_{w} \geq L\q_w \quad \forall w \in[m'],
        \end{equation}
        where $N$ is a stochastic matrix associated with the channel $\mN$, i.e. $N = [\p_1, \dots, \p_m]$.
        The matrix $S$ being stochastic $S\in \stoch(m,m')$ entails that $\s_w \geq 0$ and $\1^T_m\s_w = 1$ for all $w\in[m']$. 
        Now we phrase the existence of convex combination with coefficient $s_{x\vert w}$ as the following feasibility problem:
        \begin{align}\label{eq:primal-feasibility-majorization}
            -LN\s_{w} &\leq -L\q_w \quad \forall w \in[m'], \\
            \1_m^T\s_w &= 1 \\
            \s_w &\geq 0
        \end{align}
        Note that if there exists such a vector $\s_w$, then such vector satisfies $\1_m^T\s_w \leq 1$. Conversely, suppose there exists a vector that satisfies the problem~\eqref{eq:primal-feasibility-majorization} but, instead of equality. In that case, we have $\1_m^T\s_w \leq 1$ then rescaling $\s_w$ by a positive constant to make it equality will not violate the first condition, $-LN\s_{w} \leq -L\q_w \quad \forall w \in[m']$. Therefore, we can replace the condition $\1_n^T\s_{w}=1$ with $\1_n^T\s_{w} \leq 1$. This results in equation~\eqref{eq:primal-feasibility-majorization} being equivalent to the feasibility problem
        \begin{align}
            P\s_w &\leq \b_w \quad \forall w \in [m'] 
            \quad\text{where}\quad P \eqdef \begin{bmatrix} 
                -LN \\ \1^T _m 
            \end{bmatrix},\quad\b \eqdef \begin{bmatrix}
                -L\q_{w} \\ 1
            \end{bmatrix} \quad\text{and}\quad \s_w \geq 0
        \end{align}
        From Farkas' lemma, there exists a vector $\s_w\in\mbR^{m}_+$ such that $P\s_w\leq \b_w$ if and only if for all $\r = \t^T \oplus \lambda\in\mbR^{n+1}$,
        \begin{equation}\label{eq:farkas-imlication}
            \r^T P\geq 0 \implies \r\cdot\b_w \geq 0.
        \end{equation}
        The premise of this implication is equivalent to 
        \begin{align*}
            \lambda\1^T_m &\geq \t^T LN\\
            &= (L^T \t)^T N = (L^T \t) \cdot N.
        \end{align*}
        The least $\lambda$ satisfying this inequality is 
        \begin{equation}
            \lambda_{\text{min}} = \max_{x\in[m]} (L^T \t) \cdot \p_x.
        \end{equation}
        The latter part of the equation~\eqref{eq:farkas-imlication} is equivalent to 
        \begin{align}
            \lambda \geq (L^T\t)\cdot\q_w.
        \end{align}
        Notice that $L^T\t$ is a vector whose elements are in non-increasing order. Redefine it as $\s\in {\mbR^n_{+}}^{\da}$. Therefore, we have that $\s_w\in\mbR^{m}_+$ exists if and only if for all $\s\in\mbR^{\da,n}_{+}$ 
        \begin{equation}
            \max_{x\in[m]} \s \cdot \p_x \geq \max_{w\in[m']} \s\cdot\q_w.
        \end{equation}
\end{proof}
In proving the above lemma, we also showed that the existence of random permutation superchannel can be phrased in terms of a feasibility problem in linear programming.
\begin{corollary}
    Determining whether \(\mathcal{N} \succ \mathcal{M}\) can be efficiently solved by formulating it as a feasibility problem in linear programming.
\end{corollary}
\begin{lemma}\label{lm:operational-constructive}
    $\mN$ majorizes $\mM$ constructively if and only if $\mN$ majorizes $\mM$ operationally, i.e., the following propositions are equivalent:
    \begin{enumerate}
        \item $\mM = \Theta [\mN]$  where $\Theta$ is a random permutation superchannel.
        \item $\pr_\t (\mN) \geq \pr_\t(\mM)$ for all $\t$-games.
    \end{enumerate}
\end{lemma}
    It will be convenient to introduce the following notation:
    \begin{definition}The \emph{predictibility} function $\P_\mN:\mbb{R}^n_+\to\mbb{R}_+$ of a classical channel $\mN$ with transition matrix $[\p_1, \cdots, \p_m] \in \stoch(n,m)$ is given by
    \be
    \P_\mN(\s)\eqdef\max_{x\in[m]}\s\cdot\p_{x}^\da\;.
    \ee
    \end{definition}
\begin{proof}
    By the preceeding lemma~\ref{lm:char-predictability-random-perm}, $\mN$ majorizing $\mM$ constructively is equivalent to having
    \be\label{eq:predictability-02}
        \max_{x \in [m]} \s\cdot\p_x \geq \max_{x' \in [m']} \s\cdot\q_{x'}.
    \ee
    for any $\s\in\prob^{\da}(n)$, where $\p_x := \mN(| x\rl x|)$ for each $x \in [m]$ and $\q_{x'} := \mM(|x\rl x|)$ for each $x' \in [m']$. 

    The strategy is to show that having $\pr_\t (\mN) \geq \pr_\t(\mM)$ for all $\t$-games is equivalent to \eqref{eq:predictability-02}. First, we establish the relation between the predictability function and the winning probability of a $\t$-game. By definition of conditional probability, $t_{k\vert w} = t_{wk}/t_{\vert w}$. Applying this to Eq. (13) in the main text simplifies the expression for the winning probability to be
    \begin{align}\label{eq:prob-win-T-game-01}
    \pr_{\t}(\mN) &=\sum_{w \in[l]} \max _{x \in[m]} \sum_{k \in[n]} t_{wk}\left\|\p_x\right\|_{(k)}.
    \end{align}
    The expression inside the summation simplifies to
    \begin{align}
        \sum_{k \in[n]} t_{wk}\left\|\p_x\right\|_{(k)} &= \sum_{k \in[n]}\sum_{y \in[k]} t_{wk}\p^\da_{y\vert x}\\
        &= \sum_{y\in[n]}\sum_{k=y} ^{n} t_{kw}\p^\da_{y\vert x}.
    \end{align}
    Defining $s_{yw} \eqdef \sum_{k=y}^n t_{kw}$ and $\s_w:=\sum_{y\in[n]}s_{yw}\e_y$, we have 
    \begin{align}
        \max _{x \in[m]} \sum_{k \in[n]} t_{wk}\left\|\p_x\right\|_{(k)}  &= \max _{x \in[m]} \sum_{y\in [n]} s_{yw}\p^\da_{y\vert x} = \max _{x \in[m]} \s_w \cdot \p^\da_x = \P_\mN (s_w).
    \end{align}
    The winning probability can then be written in terms of the predictability function: 
    \begin{equation}
    \label{eq:prob-win-predictibility}
        \pr_{\t}(\mN) =\sum_{w \in[l]} \P_\mN (\s_w).
    \end{equation}
    
    Now, let us restate explicitly what we want to show:
    \be\label{eq:pr-t-implies-pred}
        \forall \t\in\bigcup_{\ell\in\mbN} \prob(n\ell) \quad \pr_\t(\mN)\geq \pr_\t(\mM)\quad \iff \quad\forall s \in \prob^\downarrow(n)\quad \max_{x\in[m]}\s\cdot\p_{x}^\downarrow \geq \max_{x'\in[m']}\s\cdot\q_{x'}^\downarrow.
    \ee
    For the left to right implication, consider $\t \in \prob^\da (n)$ and use equation~\eqref{eq:prob-win-predictibility} to get
    \begin{equation*}
        \pr _{\t}(\mN) = \P_{\mN} (\s)
    \end{equation*}
    where $\s$ is defined by $s_y = \sum_{k=y}^n t_k$. We apply the same idea to the winning chance associated with $\mM$.
    Since the relation holds for all $t\in\prob^\da(n)$, it must also hold for all $\s\in\mbR^{n,\da}_{+}$ as well.
    
    In the converse of \eqref{eq:pr-t-implies-pred}, recall that $\s_w = (s_{yw})$ is defined as 
    \begin{equation}
        s_{yw} \eqdef \sum_{k=y}^n t_{kw},
    \end{equation}
    which is not in $\prob(n)$ but is in $\mbR^{\downarrow,n}_{+}$. However, since $\left\| \s_w \right\| > 0$, the vector $\s_w/\left\| \s_w \right\|$ is in $\prob^\da (n)$ and  
    \begin{equation}
    \P_\mN (\s_w) = \max_{x\in[m]} \s_w \cdot \p_x \geq \max_{x'\in[m']} \s_w \cdot \q_{x'} = \P_\mM (\s_w).
    \end{equation}
    This leads to $\Pr_\t (\mN)\geq \Pr_\t (\mM)$, for any $\t\in\bigcup_{\ell\in\mbN} \prob(n\ell)$. This completes the proof.
\end{proof}

\begin{proof}[Proof of Theorem 4]
    From the lemma \ref{lm:cup-rand-perm}, the mixing operations defined in the constructive approach are of the same set as in those in the axiomatic approach, and by the lemma~\ref{lm:operational-constructive}, the existence of the mixing operation monotonically implies the inequality in the game of chance. 
\end{proof}

\subsection{Proof of Theorem 6}

\begin{lemma}\label{lm:predconvex}
Let $\mN, \mM$ be classical channels with transition matrices $[\p_1\cdots\p_m]\in\stoch(n,m)$,$[\p_1\cdots\p_{m-1}]\in\stoch(n,m-1)$ respectively. If there exists $\t\in\prob(m-1)$ such that 
\be
\sum_{x\in[m-1]}t_x\p_x^\da\succ\p_m
\ee 
then $\P_\mN(\s)=\P_\mM(\s)$ on all $\s \in {\mbR_{+}^n}^\da$.
\end{lemma}
    It will be convenient to introduce the following notation:
    \begin{definition}The \emph{predictibility} function $\P_\mN:\mbb{R}^n_+\to\mbb{R}_+$ of a classical channel $\mN$ with transition matrix $[\p_1, \cdots, \p_m] \in \stoch(n,m)$ is given by
    \be
    \P_\mN(\s)\eqdef\max_{x\in[m]}\s\cdot\p_{x}^\da\;.
    \ee
    \end{definition}
\begin{proof}
The predictability function can be expressed as
\be
\P_\mN(\s)=\max_{x \in [m]}\s\cdot\p_{x}^\da=\max_{\t\in\prob(m)}\sum_{x\in[m]}t_x\s\cdot\p_{x}^\da
\ee
for each $\s \in \mbb{R}^n_+$, since the maximum of the convex hull of $\{\s\cdot\p_{x}^\da\}_{x \in [m]}$ occurs on one of the $\s\cdot\p_{x}^\da$. For any $\p, \q \in \prob(n)$, if $\p \succ \q$ then
\ba
\s\cdot\p^\da = L^T\r\cdot\p^\da = \r\cdot L\p^\da = \sum_{k \in [n]}r_k\Vert\p\Vert_{(k)} \geq \sum_{k \in [n]}r_k\Vert\q\Vert_{(k)} = \r\cdot L\q^\da = L^T\r\cdot\q^\da = \s\cdot\q^\da
\ea
for any $\s \in {\mbR_{+}^n}^\da$, where $\r$ is given by $r_j=s_j-s_{j+1}$ for each $j \in [n-1]$ and $r_n = s_n$, which implies that $r_k \geq 0$ for all $k \in [n]$, from which the inequality follows. So from the hypothesis, we have
\be
\max_{\t'\in\prob(m-1)}\sum_{x\in[m-1]}t_x'\s\cdot\p_{x}^\da \geq \s\cdot\sum_{x\in[m-1]}t_x\p_{x}^\da \geq \s\cdot\p_{m}^\da
\ee
for each $\s \in {\mbR_{+}^n}^\da$. It follows that
\ba
\P_\mN(\s) &= \max_{\t \in \prob(m)} \sum_{x\in[m]}t_x\s\cdot\p_x^\da \\
&= \max_{\t \in \prob(m)} \sum_{x\in[m-1]}t_x\s\cdot\p_x^\da+t_m\s\cdot\p_m^\da \\
&= \max_{\t \in \prob(m)} (1-t_m)\sum_{x\in[m-1]}\frac{t_x}{1-t_m}\s\cdot\p_x^\da+t_m\s\cdot\p_m^\da \\
&= \max_{\lambda \in [0,1]} \max_{\t' \in \prob(m-1)} (1-\lambda)\sum_{x\in[m-1]}t_x'\s\cdot\p_x^\da+\lambda\s\cdot\p_m^\da \\
&= \max_{\t' \in \prob(m-1)} \sum_{x\in[m-1]}t_x'\s\cdot\p_x^\da \\
&= \P_\mM(\s).
\ea
for each $\s \in {\mbR_{+}^n}^\da$, where the second to last equality follows since the maximization over all convex combinations of two real numbers is simply given by the larger of those two numbers. Hence $\P_\mN = \P_\mM$ on all real vectors of non-negative, non-increasing entries.
\end{proof}

\begin{lemma}
Every channel is equivalent to its standard form under channel majorization.
\end{lemma}

\begin{proof}
Let $\mN \in \cptp(X \to Y)$ have a transition matrix $N:=[\p_1 \cdots \p_m] \in \stoch(n,m)$, and recall the predictability function
\be
\P_\mN(\s) = \max_{x \in [m]} \s\cdot\p^\da.
\ee
It follows from Theorem \ref{th:tfae-characterization} that the two channels $\mM, \mN$ are equivalent under channel majorization if $\P_\mM(\s)=\P_\mN(\s)$ for all $\s \in \prob^\da(n)$.
\ben
\item $\P_\mN$ is invariant under permutations of the components of each $\p_x$ for any choice of $x \in [m]$, and therefore under replacement of each $\p_x$ with $\p_x^\da$.
\item By Lemma \ref{lm:predconvex}, $\P_\mN$ is invariant under removal of columns of $N$ that are majorized by convex combinations of other columns of $N$. 
\item $\P_\mN$ is invariant under permutations of columns $N$.
\een
Therefore $\mN$ is equivalent to its standard form.
\end{proof}

\begin{thm}{6}
Let $\mN^{X \to Y}$, $\mM^{X' \to Y}$ be classical channels in standard form. Then $\mN^{X\to Y} \sim \mM^{X'\to Y}$ if and only if $\mN^{X\to Y}=\mM^{X'\to Y}$ (in particular, $X \cong X'$).
\end{thm}

\begin{proof}
Let $N\eqdef[\p_1\cdots\p_m]\in\stoch(n,m)$ and $M\eqdef[\q_1\cdots\q_{m'}]\in\stoch(n,m')$ be the transition matrices of $\mN$ and $\mM$, respectively. From the condition $\mN^{X \to Y}\sim\mM^{X' \to Y}$ it follows that $\mN \succ \mM$ and $\mM \succ \mN$, so from Theorem \ref{th:tfae-characterization} there exists an $m\times m'$ stochastic matrix $S=(s_{x|w})$ such that
\be
\sum_{x\in[m]}s_{x|w}\p_{x}\succ \q_{w}\quad\quad\forall\;w\in[m']\;
\ee
and an $m' \times m$ stochastic matrix $T=(t_{w|x})$ such that
\be
\sum_{w\in[m']}t_{w|x}\q_{w}\succ \p_{x}\quad\quad\forall\;x\in[m]\;.
\ee
Put in another way, there exists a stochastic matrix $S\in\stoch(m,m')$ such that $LNS\geq LM$, and a stochastic matrix $T\in\stoch(m',m)$ such that $LMT\geq LN$. Combining the two conditions together we get that the matrix $R\eqdef ST\in\stoch(m,m)$ satisfies
\be
LNR=LNST\geq LMT\geq LN\;.
\ee
Similarly, the matrix $R'\eqdef TS\in\stoch(m',m')$ satisfies $LMR'\geq LM$. The condition above implies that $LN\r_1\geq L\p_1$ which can be written as
\be
\sum_{x=2}^mr_{x|1}L\p_x\geq (1-r_{1|1})L\p_1\;.
\ee
Therefore, if $r_{1|1}<1$ then we get that a convex combination of $\p_2,\ldots,\p_m$ majorizes $\p_1$ in contradiction with the assumption that $\mN$ is given in its standard form. We therefore conclude that $r_{1|1}=1$. Similarly, for all $x\in[m]$, we must have $r_{x|x}=1$ and since $R$ is column stochastic we conclude that $R=I_m$. Using the same argument for $R'$ we conclude that also $R'=I_{m'}$. Hence, $m=m'$ and the stochastic matrices $S$ and $T$ satisfies 
\be
ST=TS=I_m\;.
\ee
That is, $T=S^{-1}$. Since the only stochastic matrix whose inverse is also a stochastic matrix is a permutation matrix we conclude that $N$ and $M$ can only differ by a permutation of their columns but since they are given in their standard form this implies that $N=M$.
\end{proof}

\subsection{Expression for the optimal upper bound of probability vectors}
Given a set $S$ of probability vectors of the same dimension, its optimal upper bound with respect to vector majorization can be computed as follows \cite{BBH+2019}: 

\begin{lemma}\label{lm:optimalupperbound}
The optimal upper bound of $S \subset \prob(n)$ in $\prob^\da(n)$ is given by $\w$ with components
\be
w_k = \frac{\sup\limits_{\p \in S} \Vert \p \Vert_{(k_j)} - \sup\limits_{\p \in S} \Vert \p \Vert_{(k_{j-1})}}{k_j-k_{j-1}}
\ee
for each positive integer $k_{j-1} < k \leq k_j$, where the $\{k_j\}$ are defined inductively by $k_0 \eqdef 0$ and
\begin{equation}
k_j \eqdef \max\left(\underset{\substack{k_{j-1}<\ell\leq n \\ \ell \in \mbN}}{\operatorname{argmax}}\frac{\sup\limits_{\p \in S} \Vert \p \Vert_{(\ell)} - \sup\limits_{\p \in S} \Vert \p \Vert_{(k_{j-1})}}{\ell-k_{j-1}}\right).
\end{equation}
for each $j \in [J]$ where $J$ is the positive integer for which $k_J=n$.
\end{lemma}

\subsection{Proof of Theorem 8}

\begin{thm}{8}\label{th:tfae-characterization}
    The following are equivalent: 
    \ben
    \item $\mN\succ\mM$.
    \item $\conv(\mN) \succ \conv(\mM)$.
    \item There exists an $m\times m'$ stochastic matrix $S=(s_{x|w})$ such that
    \be
    \sum_{x\in[m]}s_{x|w}\p_{x}\succ \q_{w}\quad\quad\forall\;w\in[m']\;.
    \ee
    \item For all $\s\in\prob^\da(n)$
    \be
    \max_{x\in[m]}\s\cdot\p_{x} ^\da \geq \max_{w\in[m']}\s\cdot\q_{w}^\da \;.
    \ee
    \item For all $\t\in\prob(n)$ and all $w\in[m']$
    \be
    \max_{x\in[m]}\sum_{k\in[n]}t_k\frac{\|\p_x\|_{(k)}}{\|\q_w\|_{(k)}}\geq 1\;.
    \ee
    \een
\end{thm}
\textit{Remark.} 5. is not in the main text.
\\
\begin{proof}
    $1. \implies 3.$ is shown in Lemma~\ref{lm:char-predictability-random-perm} 

    $3. \implies 2$. $\conv(\mN) \succ\conv(\mM)$ if for all $\q\in \conv(\mM)$ there exists $p\in\conv(\mN)$ such that $\p \succ \q$.
     Any $\q\in\conv(\mM)$ is a linear combination of $\q_y$, 
    \begin{equation}
        \q = \sum_{w\in[m']} c(w) \q_w, \quad c(w)\geq 0, \quad \sum_{w\in[m']} c(w) = 1.
    \end{equation}
    In the premises, for all $w\in [m']$ there exists a linear combination of $\p_x$ such that $\sum_{x} s(x\vert w) \p_x \succ \q_w$.
    In terms of a doubly stochastic matrix, this means 
    \begin{equation}
        \q_w = D_w \sum_{x\in[m]} s(x\vert w) \p_x.
    \end{equation}
    $\q \in \conv(\mN)$ because a doubly-stochastic can be written as a convex combination of permutation matrices. Then, for any $\q\in\conv(\mM)$, we have that $\q = \sum_{w\in[m']} c(w) \q_w$
    is also in $\conv(\mN)$. Therefore, any $\q\in \conv(\mM)$ has $
    \p\in\conv(\mN)$ which is itself $\p =\q$ and trivially $\p=\q\succ\q$.

    $2. \implies 4.$ 
    From $\conv(\mN)\succ\conv(\mM)$ we have that, for any $\q_w$ there exists $\p\in\conv(\mN)$ such that $\p\succ\q_w$. Since $\p$ is in the convex hull $\conv(\mN)$, $\p=\sum_{x\in[m]} s(x\vert w) \p_x $. Notice that $\sum_{x\in[m]} s(x\vert w) = 1$ and $s(x\vert w)\geq 0$. Then, $S = (s(x\vert w))$ is a stochastic matrix. 
    We have that $S \in \stoch(m,m')$ such that 
    \begin{equation}\label{eq:conv-majorization-vector}
        \sum_{x} s_{x\vert w} \p_x \succ \q_{w} \quad \forall w \in[m']
    \end{equation}
    coincidentally match with condition $3.$ and $3.\implies 4.$ is shown in the Lemma~$\ref{lm:char-predictability-random-perm}$.

    $4. \implies 1.$
    For all $\s\in\mbR^{\da,n}_{+}$ 
    $\max_{x\in[m]} \s \cdot \p_x \geq \max_{w\in[m']} \s\cdot\q_w$. This implies that there exists $S=(s_{x\vert w})\in\stoch(m,m')$ such that $\sum_{x\in[m']} s(x\vert w) \p_{x} \succ \q_w$ for all $w\in[m']$. This implies the existence of $D_w$ such that 
    \begin{equation}
        D_w \sum_{x\in[m]} s(x\vert w) \p_x = \q_w.
    \end{equation}
    Recall the definition of $\p_x$ and $\q_w$, define $\mD_{xw}$ to have a corresponding matrix to be $D_w$, and define that of $\mS$ to be $S$. The above equation becomes
    \begin{align}
         \mM(\e_w) &=\sum_{x\in[m]} s(x\vert w) \mD_{xw} \mN(\e_x) \\
        &= \sum_{x\in[m]} \mD_{xw} \mN(\e_x) \mS(\e_w).
    \end{align}
    This holds for all $w\in[m']$ and $\D_{xw}$ are doubly stochastic. Therefore, there exists a random permutation superchannel $\Theta$ such that $\mN= \Theta[\mM]$. 

$3.\iff 5.$ Suppose that $M_w$ is a matrix containing a ratio of Ky Fan norm between $\p_x$ and $\q_w$,
\be
    M_w := \sum_{x,k} \frac{\left\| \p_x \right\|_{(k)}}{\left\| \q_w \right\|_{(k)}} \e_x \e_k^T.
\ee
Propositions 3. and 5. can be paraphrased as 
\begin{align}
    &\text{3. There exists } \s_w\in\prob(m) \text{ such that } \min_{k\in[n]}~\s_w^T M_w \e_k \geq 1\\
    &\text{5. Every } \t \in\prob(n) \text{ satisfies } \max_{x\in[m]}~\e^T_x M_w \t \geq 1.
\end{align}
The following propositions are equivalent to proposition 3:
\begin{align}
    \exists \s_w\in\prob(m): \min_{k\in[n]}~\s_w^T M \e_k \geq 1 &\iff \min_{\t\in\prob(n)} \s^T_w M_w \t \geq 1 \\
    &\iff \max_{\s\in\prob(m)} \min_{\t\in\prob(n)} \s^T M_w \t \geq 1.
\end{align}
The following propositions are equivalent to proposition 5:
\begin{align}
    \forall \t \in\prob(n): \max_{x\in[m]}~\e^T_x M_w \t \geq 1 
    &\iff \min_{\t\in\prob(n)} \max_{x\in[m]} \e^T_x M_w \t \geq 1 \\
    &\iff \min_{\t\in\prob(n)} \max_{\s\in\prob(m)} \s^T M_w\t\geq 1
\end{align}
By the minimax theorem, the maximum and minimum can swap places, so the two propositions coincide.
\end{proof}
\subsection{Proof of Eq.~(33)}
For all systems $A$ and for all $\mN\in\cptp(A\to B)$ we have 
$$
\id^{B}\succ\mN^{A\to B} \succ \mR^{B}\;.
$$
\begin{proof}
    The majorization $\mN^{A\to B} \succ \mR^{B}$ is trivial since the $\mR^B$-replacement superchannel is mixing.

    For the other part, let $\mN^{A\to B} = \tr_R \circ \mV^{A \to RB}$ be a Stinespring representation of $\mN$. The superchannel
    \begin{equation*}
        \Theta[\mM] = \tr_R \circ \mM^{B\to B} \circ \mV^{A\to RB}
    \end{equation*}
    is mixing (for example, since tracing out $R$ is a conditionally unital operation), and takes $\id^B$ to $\mN ^{A\to B}$.
\end{proof}

\subsection{Proof of Theorem 11}
\begin{thm}{11}\label{mixing-equivalency}
    The following are equivalent for a quantum superchannel $\Theta$ with the standard form:
    \begin{equation}\label{theta-appendix}
        \Theta[\mN] = \mE^{RB \to B}\circ \mN^{A\to B} \circ {\mV^{A' \to RB}}\;.
    \end{equation}
\begin{enumerate}
    \item $\Theta$ is completely uniformity preserving
    \item The channel $\Theta[\u^B \otimes \id ^{A}]\in \cptp(A' \to AB)$ is marginally uniform with respect to $B$.
    \item $\Theta$ is mixing.
\end{enumerate}
\end{thm}
\begin{proof}
Characterization (2) follows directly from the property of complete uniformity since $\u^B \otimes \id ^{A}$ is marginally uniform.
To prove the implication $(2 \Rightarrow 3)$ we use Choi's isomorphism to translate the assumption that
\begin{equation*}
    \begin{aligned}
        \Theta\left[\u^B \otimes \id ^{A}\right] =
        \mE^{RB \to B} \circ \left( \u^B \otimes \id ^{A}\right)\circ \mV^{A' \to RA}=\mE^{RB \to B} \circ \left( \u^B \otimes \mV^{A' \to RA}\right)
    \end{aligned}
\end{equation*}
is marginally uniform to the condition:
\begin{equation*}
   \mE_{\u^B}^{R\to B} \circ \mV^{T_A} = \mR^{R\to B} \circ \mV^{T_A} \;,
\end{equation*}
where $\mE_{\u^B}^{R\to B}(\omega^R) \coloneqq \mE^{RB \to B}(\omega^R \otimes \u^B)$ and $\mV^{T_A}$ the partial transpose of $\mV$:
$\mV^{T_{A}}(\rho^{A} \otimes \omega^{A'}) \coloneqq \tr_{A} \left[\rho^T \mV(\omega)\right]$.
Because the realizatio in eq.~\ref{theta-appendix} is minimal, $\mV^{T_A}$ maps operators on the composite system $AA'$ to operators on the reference system $R$ surjectively \cite{Gour2019} and could therefore be eliminated from both sides of the equation.
 
 The implication $(3 \Rightarrow 1)$ is almost trivial:
 Given a marginally uniform channel $\mN^{AC_0 \to BC_1} = \u^B \otimes \mN ^{AC_0 \to C_1}$ we have
\begin{equation*}
\begin{aligned}
\Theta \left[ \mN^{AC_0 \to BC_1} \right] & =  
    \mE^{R\to B}_{\u^{B}}\circ  \mN^{AC_0\to C_1} \circ \mV^{A'\to R A} \\ &=
    \u^{B}\otimes  \left(\mN^{AC_0\to C_1} \circ \tr_R \circ \mV^{A'\to R A} \right)\;,
\end{aligned}
\end{equation*} 
where the second inequality follows from the conditional unitality of $\mE$. 
\end{proof}
\subsection{Proof of Theorem 13}

\begin{thm}{13}\label{paat} 
Let $\psi$ be some pure state and $\mV\in\cptp(A\to B)$ an isometry channel, then 
\be
\mV^{A\to B} \otimes \u^{ A}\sim \psi\;.
\ee
\end{thm}
We will prove a stronger version of Thm. \ref{paat} that implies it directly and also demonstrates the maximality of isometry channels with respect to majorization as follows.
\begin{lemma}\label{isometry-maximal}
Let $A,B,C,D,B',D'$ be systems such that $|BB'| = |DD'| $ and  $|AB| \ge |CD|$.
In addition, let $\mV \in \cptp(A\to B)$ be an isometry channel and $\mM\in \cptp(C\to D)$ be any channel. Then
    $$\mV^{A\to B} \otimes \u ^{B'}\succ \mM^{C\to D} \otimes \u ^{D'} \;.$$
\end{lemma}
If we look at the case $|AB|=|CD|$ and substitute $\mM$ with the isometry channel $\mU$ we get an equivalence by symmetry.
If in addition $C=\mathbb C$ (so that $|D| = |AB|$) and $B'= A$ (so that $|D'| = 1$), we get Thm.~\ref{paat}.
\begin{proof}
By the appendix version of Thm.~\ref{mixing-equivalency}
we need to find a superchannel that sends $\mV^{A\to B} \otimes \u ^{B'}$ to $\mM^{C\to D} \otimes \u ^{D'} $  and $\u^{BB'} \otimes \id^{A \to R}$ to some marginally uniform channel with respect to $BD$.
However, it is enough to find a superchannel $\Theta$ that maps $\mV^{A \to B}$ to $\mM^{C\to D}$ and $\u^{B} \otimes \id^{A \to R}$ to some marginally uniform channel with respect to $D$, since we can add and discard uniform states without changing the form of the input and the marginally uniformity of the output. Let $m$ and $n$ be the dimensions of $A$ and $B$, respictively, and abbreviate $\mR\coloneqq \mR^{A\to B}$ and $\mR' \coloneqq \mR^{C\to D}$
    consider the following supermap:
    \begin{equation*}
        \Theta[\mN^{A\to B}] \coloneqq
        \tr_{AB} \left[ \frac 1 {m^2}  J_{\mV} J_{\mN} \right]\
        \otimes \mM
        + \tr_{AB} \left[ \left( \frac n m J_{\mR} - \frac 1 {m^2}  J_{\mV} \right) J_{\mN} \right]\
          \otimes \frac{ mn \mR' -\mM}{mn-1}\;.
    \end{equation*}
    Here, 
    $$J_{\mN} \coloneqq \sum_{x,y\le m} \ket x  \bra y ^A \mN^{\tilde A \to B} \big(\ket x \bra y ^{\tilde A} \big)$$
    is the Choi matrix of a channel $\mN \in \cptp({A\to B})$, and $\{\ket x _A\}_{x \le m}$ denotes a basis of $A$.

Given $\mN\in \cptp(A\to B)$ note that 
\begin{equation*}
    0 \le \frac 1 {m^2} \tr_{AB} \left[  J_\mV J_{\mN} \right] 
    \le 1
\end{equation*}
and
\begin{equation*}\label{inner-product-with-mixing}
     \frac {n}{m}\tr_{AB} \left[  J_{\mR} J_{\mN} \right] =  
     \frac {n}{m}\tr_{AB}\left[\1^A \otimes \u^B J_{\mN}\right]
     =\frac 1 m \tr_{AB}\left[J_{\mN}\right] = 1
\end{equation*}
where in the first equality we have used the fact that that the Choi-matrix of $\mR = \u^B\circ \tr_A$ is $\1^A \otimes \u^B$.
Therefore
$\tr_{AB} \left[ \frac 1 {m^2}  J_{\mV} J_{\mN} \right]$ and $\tr_{AB} \left[ \left( \frac n m J_{\mR} - \frac 1 {m^2}  J_{\mV} \right) J_{\mN} \right]$ form a probability vector for every input channel $\mN$. 

$|CD| \mR'- \mM$ is completely positive, so by the assumption that $mn \ge |CD|$, the map $$\frac{ mn \mR' -\mM}{mn-1}$$ is a quantum channel. We conclude that $\Theta$ is a measure-prepare superchannel.

Moreover, since $\mV$ is an isometry channel, $J_\mV$ is a pure (unnormalized) state, hence
$\frac 1 {m^2} \tr_{AB} \left[  J_\mV ^2 \right] 
    = 1$, which means $\Theta[\mV] = \mM$.
    
To complete the proof, it is left to show that $\Theta[\u^B \otimes \id^{A\to R}]$ is marginally uniform with respect to $D$.
Note that for any two channels, $\mE,\mN\in \cptp(A\to B)$, we have
    \begin{equation*}
        \tr_{AB} \left[ J_{\mN} J_{\mE} \right] = 
        \sum_{x,y\le |A|}\tr_B \left[ \mE(\ket x \bra y_A)^* \mN(\ket x \bra y_A)\right]\;,
    \end{equation*}
and therefore, for every channel $\mN$
\begin{equation*}
\begin{aligned}
\tr_{AB} \left[ J_{\mN} J_{\big(\u^B \otimes \id^{A\to R}\big)} \right]
     & = \sum_{x,y\le |A|}\tr_B \left[ \mN(\ket x \bra y_A)^* \left(\u^B \otimes \id^{A\to R}(\ket x \bra y_A)\right)\right] 
    \\ & =
    \sum_{x,y\le |A|}\tr_B \left[ \mN(\ket x \bra y_A)^* \u^B \right] \ket x \bra y _R \\ &
    =\frac 1 n\sum_{x,y\le |A|}\tr_B \left[ \mN(\ket x \bra y_A)^*  \right] \ket x \bra y _R \\ &
    = \frac 1 n\sum_{x,y\le |A|}\delta_{xy} \ket x \bra y _R = \frac mn \u^R\;,
\end{aligned}
\end{equation*}
where in the fourth equality we have used the fact that $\mN$ is trace preserving.
Substituting this in the definition of $\Theta$ amounts to:
\begin{equation*}
\begin{aligned}
    \Theta\big[\u^B \otimes \id^{A\to R}\big] & = \frac 1 {mn} \mM \otimes \u ^R + \left(1 - \frac 1 {mn}\right) \frac{ mn \mR' -\mM}{mn-1} \otimes \u^R
    = \mR' \otimes \u^R \;.
\end{aligned}
\end{equation*}
\end{proof}
As a corollary, we get the following
\begin{corollary}\label{corollary}
Let $A,B,C,D,B',D'$ be systems such that $|BB'| = |DD'| $ and  $|AB| > |CD|$.
In addition, let $\mV \in \cptp(A\to B)$ be an isometry channel, then 
    \begin{equation*}
        \mV^{A\to B} \otimes \u ^{ B'}\succnsim \mM^{C\to D} \otimes \u ^{D'}
    \end{equation*}
    for every channel $\mM \in \cptp(C\to D)$.
\end{corollary}
\begin{proof}
Assume to the contrary. First note that by adding a pure state $\psi^C$ to both sides of the equation, we can assume without loss of generality that  $|C|\le|D|$.
Now, by Lemma \ref{isometry-maximal} we have $\mU^{C\to D} \succ \mM^{C\to D}$ for some isometry channel $\mU$, and by transitivity of majorization we must have
$$ \mV^{A\to B} \otimes \u ^{ B'}\sim \mU^{C\to D} \otimes \u ^{D'}\;.$$
By adding $\u^{AC}$ to both sides of the equation we get
$$\mV^{A\to B} \otimes\u^A \otimes \u ^{ CB'}\sim \mU^{C\to D}\otimes \u^C \otimes \u ^{AD'}\;.
$$
By Thm.~\ref{paat} the right hand side is equivalent to $\u^{B'C}$ and the left hand side to $\u^{AD'}$, but those are uniform states and channel  majorization between them reduces to state majorization, so we arrive to contradiction since $|AD'|>|B'C|$.
\end{proof}

\subsection{Proof of Theorem 14}

\begin{thm}{14}\label{th:condmix}
The channel $\mN^{A'B\to AB}$ is a conditionally mixing operation if and only if it has the form
\be\label{semi-appendix}
\mN^{A'B\to AB}=\mE^{RB\to B} \circ \mV^{A' \to RA}\;,
\ee
where $\mV^{A' \to RA}$ is an isometry and $\mE^{RB\to B}$ is a conditionally unital channel. 
\end{thm}

\begin{proof}
Suppose first that $\mE_\tau$ is doubly stochastic for every density matrix $\tau^R$. Then, since every Hermitian matrix $\eta^R$ can be expressed as $\eta^R=a\tau_1^R-b\tau_2^R$, where $a,b\geq 0$ are non negative real numbers and $\tau_1$ and $\tau_2$ are two density matrices, we get from linearity that 
$\mE_\eta^{A\to A}(\cdot)\eqdef\mE^{RA\to A}\left(\eta^R\otimes(\cdot)\right)$ satisfies
\be\label{40}
\mE_\eta^{A\to A}(\u^A)=\tr\left[\eta^R\right]\u^A\;.
\ee
Now, let $\rho^B$ be a state and denote by  
\be
\omega^{RB'}\eqdef\mV^{B\to RB'}\left(\rho^B\right)\;.
\ee
Thus, from~\eqref{semi-appendix} we get
\be
\mN^{AB\to AB'}\left(\u^A\otimes\rho^B\right)=\mE^{RA\to A}\left(\u^A\otimes\omega^{RB'}\right)\;.
\ee
Decomposing $\omega^{RB'}=\sum_{j\in[k]}\eta^R_j\otimes\omega_j^{B'}$ with $\eta_j$ and $\omega_j$ being hermitian matrices, we conclude that
\ba
\mN^{AB\to AB'}\left(\u^A\otimes\rho^B\right)&=\sum_{j\in[k]}\mE^{RA\to A}\left(\u^A\otimes\eta_j^{R}\right)\otimes\omega_j^{B'}\\
\eqref{40}\rightarrow&=\u^A\otimes\sum_{j\in[k]}\tr\left[\eta_j^R\right]\omega_j^{B'}\\
&=\u^A\otimes\omega^{B'}\;.
\ea
Thus, $\mN^{AB\to AB'}$ is conditional unital.

Conversely, suppose $\mN^{AB\to AB'}$ is a conditionally mixing operation. Since $\mN^{AB\to \tA B'}$ is conditionally unital, one of the marginals of its Choi matrix satisfies (see Lemma 7.1.1. in~\cite{Gour2024})
\be\label{44}
J_{\mN}^{B\tA B'}=J_\mN^{BB'}\otimes\u^{\tA}\;.
\ee
Since $\mN$ is also $A\not\to B'$ signalling, we get from~\eqref{semi-appendix} that
\be
J_{\mN}^{B\tA B'}=|AB|\mE^{RA\to \tA}\left(\u^A\otimes\phi^{BB'R}\right)
\ee
where
\be
\phi^{BB'R}=\mV^{\tB\to RB'}\left(\Phi^{B\tB}\right)
\ee
and $\Phi^{B\tB}$ is a maximally entangled state. Combining this with~\eqref{44} and denoting by $\sigma^{BB'}\eqdef\tr_r\left[\phi^{BB'R}\right]$ gives
\be
\mE^{RA\to A}\left(\u^A\otimes\phi^{BB'R}\right)=\u^{A}\otimes\sigma^{BB'}\;.
\ee
Since $R$ has the same dimension as the support of $\sigma^{BB'}$ we can embed $\sigma$ in $R$. Thus, conjugating both sides of the equation above by $\sigma^{-1/2}(\cdot)\sigma^{-1/2}$ gives
 \be
\mE^{RA\to A}\left(\u^A\otimes\Omega^{\tR R}\right)=\u^{A}\otimes I^{\tR}\;,
\ee
where we restricted $\sigma^{BB'}$ to its support $\tR\eqdef\supp(\sigma^{BB'})$. Finally, multiplying both sides by a density matrix $(\tau^{\tR})^T$ and taking the traces gives
\be
\mE^{RA\to A}\left(\u^A\otimes\tau^{R}\right)=\u^{A}\;.
\ee
Since $\tau^R$ was arbitrary, this completes the proof.
\end{proof}
\subsection{Proof of Theorem 15}

\begin{thm}{15}\label{th:conditional-channel}
Let $\mN^{X \to Y}$ and $\mM^{X' \to Y}$ be two classical channels. Then $\mN^{X\to Y} \succ \mM^{X' \to Y}$ if and only if 
\be
\sum\limits_{x \in [m]} p_x\e_x^X \otimes \mN^{X \to Y}(\e_x^X) \succ_Y \sum\limits_{w \in [m']} q_w \e_w^{X'} \otimes \mM^{X'\to Y} (\e_w^{X'})
\ee
for some $\mathbf{p}\in \prob(m)$ and $\mathbf{q}\in \prob(m')$.
\end{thm}

\begin{proof}
     Denote $|X|=m,\ |X'|=m',\ |Y|=n$, $\p_x^Y = \mN(\e_x^X)$, $\q_w ^Y = \mM(\e_w ^{X'})$,  and assume $\sum_x p_x\e_x \otimes \p_x \succ_Y \sum_w q_w \e_w \otimes \q_w $. The latter is equivalent to $\sum_{xy}p_xp_{y \vert x}\e_x \otimes \e_y \succ_Y \sum_{wy}q_{w}q_{y \vert w}\e_{w} \otimes \e_y$ for transition matrices $N=(p_{y \vert x}), M=(q_{y \vert x'})$ of $\mN,\mM$ respectively. Writing $\p_x = \sum_y p_{y \vert x}\e_y$ and $\q_w =\sum_y q_{y \vert w}\e_y$, it follows from Theorem 4.6.2. of \cite{Gour2024} that this is equivalent to the existence of $R=(r_{w|x})\in\text{STOCH}(m',m)$ such that 
    \begin{align}
     \sum_{x\in[m]} r_{w|x}p_x\textbf{p}_{x}^{Y} \succ q_{w}\textbf{q}_{w}^{Y} \qquad \forall w\in[m'],
    \end{align}
    or equivalently
    \begin{align}\label{eq:1}
     \sum_{x\in[m]}\frac{r_{w|x}p_x}{q_w}\textbf{p}_{x}^Y \succ \textbf{q}_{w}^Y\qquad \forall w\in[m'].
    \end{align}
     The LHS of \eqref{eq:1} is a convex combination (as can be seen by taking the sum of all elements on both sides), and so $\mN^{X \to Y} \succ \mM^{X' \to Y}$ according to Theorem \ref{th:tfae-characterization}.
 
     For the other direction, we will do the reverse process. Assume $\mN \succ \mM$. Then according to Theorem \ref{th:tfae-characterization} there exists $\Tilde{R}=(\Tilde{r}_{x|w})\in\text{STOCH}(m,m')$ such that
     \begin{equation}
     \label{eq:2}\sum_{x\in[m]}\Tilde{r}_{x|w}\textbf{p}_{x}^Y\succ \textbf{q}_{w}^Y\qquad \forall w\in[m'].
     \end{equation}
     Choose any $\mathbf{q}\in \prob(m')$ and define $\mathbf{p} \in \prob(m)$ according to
     \begin{equation}
         p_x:=\sum_{w\in[m']}\Tilde{r}_{x|w}{q_{w}}\qquad \forall x\in[m].
     \end{equation}
    Define $S=(s_{w|x})\in\text{STOCH}(m',m)$ via $s_{w|x}:=\Tilde{r}_{x|w}\frac{q_{w}}{p_x}$ and from \eqref{eq:2} get 
    \begin{align}
    \sum_{x\in[m]}
    \frac{s_{w|x}p_x}{q_{w}}\textbf{p}_{x}^Y\succ \textbf{q}_{w}^Y\qquad \forall w\in[m']
    \end{align}
    or
    \begin{align}
    \sum_{x\in[m]}
    s_{w|x}p_x\textbf{p}_{x}^Y\succ q_w \textbf{q}_{w}^Y\qquad \forall w\in[m']
    \end{align}
    which implies
    \be
    \sum_{x\in[m]} p_x\e_x^{X} \otimes \p_x^{Y} \succ_Y \sum_{w\in[m']} q_w \e_w^{X} \otimes \q_w^{Y},
    \ee
    finishing the proof.
\end{proof}

\subsection{Normalization of a Channel Entropy}
Hereafter, whenever the system $A$ is 1-dimensional, and hence $\cptp(A\to B)$ is identified with the set of quantum states on $B$ by $\mN \mapsto \mN(1)$, we denote $\H(\rho) \coloneqq \H(A|B)_\rho$.
\begin{lemma}
    Let $\H$ be an unnormalized non-zero entropy of channels and denote the uniform qubit state by $\u_2$. Then $\H(\u_2) > 0$.
\end{lemma}
The proof follows the same lines as in ref.~\cite{GWBG2022}, we provide it here for completeness.
\begin{proof}
Note that $\u_2 \succ \u_2^{\otimes 2}$ (since $\u_2$ can be embeded in $\mathbb C^4$ where $\u_2^{\otimes 2}$ is minimal) and therefore $\H(\u_2) \le 2 \H(\u_2)$ so $\H(\u_2) \ge 0$. We need to show that this inequality is strict.
    By assumption, there exists a channel $\mN \in \cptp(A\to B)$ of non-zero entropy.
    Without loss of generality, using a post-processing isometry channel, we can assume $B = \mathbb C^{2^m}$ for some $m\in \mathbb N$.
    Minimality of the uniform state tells us that $\mN \succ \u^B = \u_2^{\otimes m}$ and because of additivity, if the entropy of $\mN$ is positive, so is that of $\u_2$. We therefore proceed assuming $\H(B | A)_\mN<0$. 
    
    By the additivity axiom, the state $1$ on the trivial system $\mathbb C$ must have zero entropy.
    Again by invariance under post-processing isometry channels so does any other pure state.
    Let $k \in \mathbb N$ such that $2^k> |AB|$.
    By Lemma \ref{isometry-maximal} any pure state $\psi$ on $\mathbb C^{2^k}$ satsifies $\psi \succ \mN\otimes \u^{R}$ for a system $R$ of dimension $2^k - |B|$.
    But as before, we have $\mN\otimes \u^{R} \succ \mN\otimes\u_2^{\otimes k}$, which by additivity means that
    $
        \H(\u_2) = -\H(B | A)_\mN/k>0
    $.
\end{proof}
\subsection{Equivalence between Definitions of the Min Entropy of Channels}
\begin{lemma}
    The definitions of the min-entropy of channels are equivalent
    \begin{align}
        \label{min-entropy-cp}
        H_{\rm min}\left(B|A\right)_{\mN} &:=\log|B|-D_{\max}\left(\mN\big\|\mR\right)\\
        \label{min-entropy-pure}
        H_{\min}(B|A)_{\mN} &\eqdef\min_{\psi\in\pure(RA)}H_{\min}\left(B|A\right)_{\mN^{A\to B}\left(\psi^{RA}\right)}
    \end{align} 
\end{lemma}
\begin{proof}
Note that $\mR^{A\to B}(\psi^{RA}) = \psi^R \otimes \u^B$ for all $\psi\in \pure(RA)$, hence it is enough to show that
$$D_{\max}\left(\mN\big\|\mR\right) = \max_{\psi \in \pure(RA)}D_{\max}\left(\mN^{A \to B}(\psi^{RA})\big\|\mR^{A \to B}(\psi^{RA})\right)\;. $$
The proof is almost immediate from Choi's characterization of completely positive maps.
First, note that by definition, if $t \mR - \mN $ is completely positive then $(t \mR - \mN)^{A \to B}(\psi^{AR}) \ge 0$ for every pure state $\psi^{RA}$, so we conclude
$$D_{\max}\left(\mN\big\|\mR\right) \ge \max_{\psi \in \pure(RA)}D_{\max}\left(\mN^{A \to B}(\psi^{RA})\big\|\mR^{A \to B}(\psi^{RA})\right)\;. $$
On the other hand, by Choi's characterization $(t \mR - \mN)$ is completely positive whenever its application on the maximally entangled state, $(t \mR - \mN)^{A \to B}(\Phi^{RA}) = t\u^{RB} - \mN^{A\to B}(\Phi^{RA} )$ is positive. Hence,
$$D_{\max}\left(\mN\big\|\mR\right) = D_{\max}\left(\mN^{A \to B}(\Phi^{RA})\big\|\mR^{A \to B}(\Phi^{RA})\right) \le  \max_{\psi \in \pure(RA)}D_{\max}\left(\mN^{A \to B}(\psi^{RA})\big\|\mR^{A \to B}(\psi^{RA})\right)\;. $$
\end{proof}

\subsection{Proof of Theorem 18}

\begin{thm}{18}
Let $\H$ be an entropy of a quantum channel. For all $\mN\in\cptp(A\to B)$
\be
\H(B|A)_{\mN}\geq H_{\min}(B|A)_{\mN}.
\ee
\end{thm}
\begin{proof}
  Assume to the contrary, that is, assume that there exist systems $A, B$ and a channel $\mN\in \cptp(A \to B)$ 
  such that $\H(B|A)_\mN< H_{\min}(B|A)_\mN$.
  First, we will manipulate the channel $\mN$ to translate this separation and ensure it contains a positive rational and then inflate it to include a natural number.
  
From additivity, for any $k\in \mathbb N$
   \begin{equation}
    \H(B\otimes \mathbb C^{2^k}|A)_{\mathcal N\otimes \u_2^{\otimes k}} = \H(B|A)_\mN+k < H_{\min}(B|A)_\mN+ k = H_{\min}(B\otimes \mathbb C^{2^k}|A)_{\mathcal N\otimes \u_2^{\otimes k}}\;,
   \end{equation}
   where $\mathbf{u}_2$ is the uniform qubit state.
   Specifically, because $k$ is arbitrarily large we can assume that $\H(B|A)_\mN>0$ without loss of generality (by renaming $\mN$). Therefore, there exists a positive rational number $t=\frac m n$ that satisfies $\H(B|A)_\mN<t< H_{\min}(B|A)_\mN$. 
   Furthermore, in such case, 
   \begin{equation}
    \H( B^n | A^n)_{\mathcal N^{\otimes n}} = n \H(B|A)_\mN
    <m<
     nH_{\min}(B|A)_\mN = H_{\min}( B^n | A^n)_{\mathcal N^{\otimes n}}
   \end{equation}
  so by renaming $\mN$ again, we may actually assume that $t=m\in \mathbb N$.

We shall now prove that $\u_2^{\otimes t} \succ \mathcal N$ leading to
  $t = \H\left( \u_2^{\otimes t} \right) \le \H\left(B|A \right)_\mN$ which is a contradicition. Define $l = 2^t $ and note that 
  $$\log l = t<H_{\min}\left( B|A \right)_\mN \le H_{\min}\left( \u^B \right) = \log n\;,$$
  where $n$ is the dimesion of $B$. We can therefore identify $\u_l \coloneqq \u_2^{\otimes t}$ with its image under an isometry channel in $\cptp\big(\mathbb C^l \to B\big)$ (by adding zeros) and viewing it as a state in $B$.
  Define $\Theta$ to be the supermap taking operators on $B$ to superoperators on $(A\to B)$ in the following way: 
  \begin{equation*}
  \Theta[\rho] \coloneqq \tr \left( \Lambda \rho \right) \mathcal N
  + \left( 1 - \tr\left( \Lambda \rho \right) \right) 
  \frac{n \mathcal R - l\mathcal N}{n-l}\;,
  \end{equation*}
  Where $\Lambda$ is the projection in $B$ on the support of $\u_l$ and $\mR \coloneqq \mR^{A\to B}$.
  By definition
  $\tr\left( \Lambda \u_l \right) = 1$, so 
  $\Theta$ takes $\u_l$ to $\mathcal N$, hence, we will be done once we prove that $\Theta$ is a mixing superchannel.

  First, since $\Theta$ is a measurement-prepare supermap, in order to prove that it is indeed a superchannel it is enough to show that $\frac{n \mathcal R - l\mathcal N}{n-l}$ is a quantum channel.
  It is trace preserving as an affine combination of channels. To see that it is completely positivite, observe that 
  $$l=2^t \le 2^{H_{\min}\left( B|A\right)_{\mN}} = \frac{n}{ \min\left\{s \mid s \mR \ge \mN\right\}}\;,$$
  so
  $$n \ge \min\left\{l s \mid s \mR \ge \mN\right\} = \min\left\{s \mid s \mR \ge l \mN\right\}\;,$$
which means $n\mR \ge l \mN$.
  
  Since the inputs of $\Theta$ are states, in order to check that it is mixing, we only need to check that it takes the uniform state to the uniform channel.
  Indeed, note that $\tr(\Lambda \u^B) = \frac {l} {n}$ so \begin{equation*}
    \Theta\left[ \u^B \right] = \frac {l} {n} \mathcal N + \frac {n-l} {n} \cdot\frac  {n \mathcal R - l \mathcal N}{n-l} =  \mathcal R\;.
  \end{equation*}
\end{proof}

\subsection{Proof of Theorem 19}

\begin{thm}{19}\label{th:maxext}
Let $\mN^{X \to Y}$ be a classical channel with $\p_x:=\mN(\e_x)$. The maximal extension of any quasiconcave classical entropy function $\H$ is given by
    \be
    \overline{\H}(Y|X)_\mN=\min_{x\in[m]}\H(\p_x)
    \ee
is an entropy
\end{thm}

\begin{proof}
    By definition, the maximal extension $\overline{\H}$ of classical entropy function $\H$ is given by 
    \begin{equation}
        \overline{\H}(Y \vert X)_{\mN} = \inf\{\H(\q) : \mN \succ \q \}
    \end{equation}
    where the infimum is among all probability vectors $\q$ of any dimension.
    
    A probability vector $\q$ satisfies $\mN \succ \q$ if and only if there exists a convex combination of $\p_x$ which majorizes $\q$; that is, $\sum_{x \in [m]} \lambda_x \p_x \succ \q$ for some $\lambda_x \geq 0$ such that $\sum_{x \in [m]}\lambda_x = 1$. It follows from the Schur-concavity of $\H$ that
    \begin{equation}
        \H\!\left(\sum_{x \in [m]} \lambda_x \p_x\right) \leq \H\left(\q\right).
    \end{equation}
    Since $\H$ is quasiconcave, we have
    \begin{equation}
        \min_{x \in [m]}\H(\p_x) \leq \H\!\left(\sum_{x \in [m]} \lambda_x \p_x\right) \leq \H\left(\q\right),
    \end{equation}
    which shows that $\min\{\H(\p_x): x \in [m]\}$ is a lower bound for $\{\H(\q): \mN \succ \q\}$. This lower bound is attained since $\mN \succ \p_{x_0}$, where $x_0 \in [m]$ such that $\H(\p_{x_0}) = \min\{\H(\p_x): x \in [m]\}$. This completes the proof.
\end{proof}

\begin{lemma}\label{lm:maxextadd}
The maximal extension of a quasiconcave classical entropy $\H$ is additive.
\end{lemma}

\begin{proof}
Let $\mN \in \cptp(X \to Y)$ and $\mM \in \cptp(X^\prime \to Y^\prime)$, and denote $|X|=m, |X^\prime|=m^\prime$. Then using the form for the maximal extension of a quasiconcave classical entropy function $\H$, we have
\begin{align}
\overline{\H}(YY^\prime \vert XX^\prime)_{\mN \otimes \mM} &= \min_{z \in [mm^{\prime}]}\H((\mN \otimes \mM)(|z \lr z|)) \\
&= \min_{x \in [m], x^{\prime} \in [m^{\prime}]}\H((\mN \otimes \mM)(|x \lr x| \otimes |x^{\prime} \lr x^{\prime}|)) \\
&= \min_{x \in [m], x^{\prime} \in [m^{\prime}]}\H(\mN(|x \lr x|) \otimes \mM(|x^{\prime} \lr x^{\prime}|)) \\
&= \min_{x \in [m], x^{\prime} \in [m^{\prime}]}\big[\H(\mN(|x \lr x|))+\H(\mM(|x^{\prime} \lr x^{\prime}|))\big] \\
&= \min_{x \in [m]}\H((\mN(|x \lr x|))+\min_{x^{\prime} \in [m^{\prime}]}\H(\mM(|x^{\prime} \lr x^{\prime}|)) \\
&= \overline{\H}(Y \vert X)_{\mN}+\overline{\H}(Y^\prime \vert X^\prime)_{\mM}.
\end{align}
Therefore $\overline{\H}$ is additive.
\end{proof}

\begin{lemma}
The minimal extension of the entropy function $\H$ to the domain of classical channels is given by
\be
    \underline{\H}(Y \vert X)_{\mN}=\H(\q)
\ee 
where $\q$ is the optimal upper bound of $\{\mN(|x \lr x|)\}_{x \in [|X|]}$ with respect to majorization. 
\end{lemma}

\begin{proof}
Let $\mN \in \cptp(X \to Y)$, and denote $|X|=m, |Y|=n$. The minimal extension of $\underline{\H}$ is given by
\begin{align}
\underline{\H}(Y \vert X)_{\mN} &= \sup\limits_{\p \in \!\bigcup\limits_{N \in \mbN} \!\!\prob(N)}\{\H(\p) \;\vert\; \p \succ \mN\} \\
&= \sup\limits_{\p \in \!\bigcup\limits_{N \in \mbN} \!\!\prob(N)}\{\H(\p) \;\vert\; \forall x \in [m]: \p \succ \mN(|x \lr x|)\} \\
&= \sup\limits_{\p \in \prob(n)}\{\H(\p) \;\vert\; \forall x \in [m]: \p \succ \mN(|x \lr x|)\} \\
&= \H(\q)
\end{align}
where $\q$ is the optimal upper bound of $\{\mN(|x \lr x|)\}_{x \in [m]}\}$; that is, $\q$ is the unique vector in $\prob^\da(n)$ for which $\q \succ \mN(|x \lr x|)$ for all $x \in [m]$, and $\p \succ \q$ for any $\p \succ \mN(|x \lr x|)$ for all $x \in [m]$.

The third equality follows from how if $\p \in \prob(N)$ for $N>n$, then $\p \succ N(|x \lr x|) \in \prob(n)$ for each $x \in [m]$, so
\be
1 = \Vert N(|x \lr x|) \Vert_{(n)} = \Vert N(|x \lr x|)\oplus\mathbf{0} \Vert_{(n+k)} \leq \Vert \p \Vert_{(n+k)} \leq 1
\ee
which implies $p^\da_{n+k}=0$ for any $k \in [N-n]$. Therefore it is sufficient to consider the infimum over probability vectors of dimension $n$. 

The final equality follows from $H$ being an antitone of the majorization preorder, and any collection of probability vectors of the same dimension having an optimal upper bound.

We add that the optimal upper bound $\q$ of $\{\mN(|x \lr x|)\}_{x \in [m]}$ can be computed directly from Lemma \ref{lm:optimalupperbound} as follows:
\be
    q_k = \frac{\pr_{k_j}(\mN) - \pr_{k_{j-1}}(\mN)}{k_j-k_{j-1}}
\ee
for each positive integer $k_{j-1} < k \leq k_j$, where the $\{k_j\}$ are defined inductively by $k_0 \eqdef 0$ and
\be
    k_j \eqdef \max\left(\underset{\substack{k_{j-1}<\ell\leq n \\ \ell \in \mbN}}{\operatorname{argmax}}\frac{\pr_\ell(\mN) - \pr_{k_{j-1}}(\mN)}{\ell-k_{j-1}}\right)
\ee
for each $j \in [J]$ where $J$ is the positive integer given by $k_J=n$ and
\be
    \pr_k(\mN) \eqdef \max_{x \in [m]}\Vert \mN(|x \lr x|) \Vert_{(k)}
\ee
for each $k \in [n]$.
\end{proof}

\subsection{Proof of Theorem 20}

\begin{thm}{20}\label{thm:shannonunique}
Let $\h$ be a channel entropy that reduces to the Shannon entropy of probability vectors. Then for every classical channel $\mN$,
\ba
	\h(Y|X)_\mN &= \underline{H}^\reg(Y|X)_\mN\\&=\overline{H}(Y|X)_\mN\\&=\min_{x \in [m]} H(\p_x).
\ea
\end{thm}

The proof of this theorem proceeds nearly identically to the proof of Theorem 4 in \cite{Gour2021}, while noting that the proof of Lemma \ref{lm:entropytensor} has been corrected to not rely on the incorrect formula for the optimal upper bound. Denote by
\be
    \bigvee_{x \in [m]} \a_x
\ee
the unique optimal upper bound of $\{\a_x\}_{x \in [m]} \subset \prob(n)$ in $\prob^\da(n)$. We first prove the following lemma:

\begin{lemma}\label{lm:entropytensor}
Let $\ell, m, n \in\mbb{N}$, $\p_{1},...,\p_m\in\prob(n)$, and $\s\in\prob(\ell)$. Then
\be
H\!\left(\bigvee\limits_{x \in [m]}(\p_x \otimes \s)\right)
\leq\log(\ell)+H\!\left(\bigvee\limits_{x \in [m]}\p_x\right).
\ee
\end{lemma}

\begin{proof}
Observe that since $\s \succ \u^{(\ell)}$, we have $\p_x\otimes\s\succ\p_x\otimes\u^{(\ell)}$ for any $x \in [m]$. Therefore, $\bigvee_{x \in [m]}(\p_x \otimes \s) \succ \bigvee_{x \in [m]}(\p_x \otimes \u^{(\ell)})$, from which the Schur-concavity of $H$ implies
\be
H\!\left(\bigvee\limits_{x \in [m]}(\p_x \otimes \s)\right) \leq H\!\left(\bigvee\limits_{x \in [m]}(\p_x \otimes \u^{(\ell)})\right).
\ee

\noindent Letting $\t:=\bigvee_{x \in [m]} (\p_x \otimes \u^{(\ell)})$ and $\v$ be the marginal of $\t$ obtained by $v_y = \sum_{z=1}^{\ell} t_{(y-1)\ell+z}$ for each $y \in [n]$, we have
\be
\Vert \p_x \Vert_{(\beta)} = \Vert \p_x \otimes \u^{(\ell)} \Vert_{(\beta\ell)} \leq \Vert \t \Vert_{(\beta\ell)} = \Vert \v \Vert_{(\beta)}
\ee
for each $x \in [m]$ and $\beta \in [n]$, so $\v \succ \p_x$ for each $x \in [m]$, hence $\v \succ \bigvee_{x \in [m]}\p_x$. Now, observe that
\be
(\v \otimes \u^{(\ell)})_{k} = \sum_{k'\in[n\ell]}D_{kk^{\prime}}t_{k^{\prime}}
\ee
for each $k \in [n\ell]$, where $D_{kk'}$ is a doubly stochastic matrix given by
\be
D_{(y-1)\ell+z,(y^\prime-1)\ell+z^\prime} \eqdef \frac{\delta_{yy^{\prime}}}{\ell}
\ee
for each $y, y' \in [n], z, z' \in [\ell]$. This implies that
\be
\t \succ \v \otimes \u^{(\ell)},
\ee
and it follows from the Schur-concavity of $H$ that
\be
H\!\left(\bigvee\limits_{x \in [m]}(\p_x \otimes \s)\right) =H(\t) \leq H(\v \otimes \u^{(\ell)}) = H(\u^{(\ell)})+H(\v) \leq \log(\ell)+H\!\left(\bigvee\limits_{x \in [m]}\p_x\right),
\ee
completing the proof.
\end{proof}

\noindent We are now ready to prove the main theorem.

\begin{proof}[Proof of Theorem \ref{thm:shannonunique}]
Denote by $m \eqdef |X|$ and $n\eqdef|Y|$, and for each $x\in[m]$ denote $\p_x\eqdef\mN(|x \lr x|)\in\prob(n)$. Given $x \in [m^k]$, write $x=x_1+\sum_{j \in [k-1]} (x_{j+1}-1) m^j$, where $x_j \in [m]$ for each $j \in [k]$, from which we get $|x \lr x| = \otimes_{j=1}^k |x_j \lr x_j|$. Letting $\t\eqdef(t_1,...,t_m)$ be the type of the sequence $\x\eqdef(x_1,...,x_k)\in[m]^k$, and $\mathfrak{T}_{m,k}$ the set of all types of sequences in $[m]^k$, we have
\begin{align}
\underline{H}\left(Y^k \vert X^k\right)_{\mN^{\otimes k}}&:=H\!\left(\bigvee_{x \in [m^k]} \mN^{\otimes k}(|x \lr x|)\right) = H\!\left(\bigvee_{x \in [m^k]} \mN^{\otimes k}\left(\bigotimes_{j=1}^k |x_j \lr x_j|\right)\right) \\
&= H\!\left(\bigvee_{x \in [m^k]} \bigotimes_{j=1}^k \mN(|x_j \lr x_j|)\right) =H\!\left(\bigvee_{\x \in [m]^k} \bigotimes_{j=1}^k \p_{x_j} \right)\\
&=H\!\left(\bigvee_{\t\in\mathfrak{T}_{m,k}} \bigotimes_{x=1}^m \p_{x}^{\otimes kt_x} \right)\\
&\geq -\log\left((n^{\kappa})^m\right))+H\!\left(\bigvee_{\t\in\mathfrak{T}_{m,k}} \left(\bigotimes_{x=1}^m \p_{x}^{\otimes kt_x} \otimes \bigotimes_{x=1}^m \p_{x}^{\otimes \kappa} \right) \right) \\
&= -\kappa m\log(n)+H\!\left(\bigvee_{\t\in\mathfrak{T}_{m,k}} \left(\bigotimes_{x=1}^m \p_{x}^{\otimes kt_x+\kappa} \right) \right)
\end{align}
where $\kappa$ is an integer $1\leq\kappa\leq k$, and the inequality follows from Lemma~\ref{lm:entropytensor}.

Following the same line of reasoning as the proof of Theorem 4 in the supplemental material section of \cite{Gour2021} shows that
\be
\underline{H}^{\reg}(Y \vert X)_{\mN} \eqdef \lim_{k \to \infty}\frac{1}{k}\underline{H}\left(Y^k \vert X^k\right)_{\mN^{\otimes k}} \geq \min_{x\in[m]}H(\p_x) = \overline{H}(Y \vert X)_{\mN}
\ee
where the last equality follows from Theorem \ref{th:maxext} and the quasiconcavity of the Shannon entropy. Combining this with the properties of regularization and the expression for the maximal extension of $\H$ we get
\be
\overline{H}^{\reg}(Y \vert X)_{\mN}=\overline{H}(Y \vert X)_{\mN} \leq \underline{H}^{\reg}(Y \vert X)_{\mN} \leq \H(Y \vert X)_{\mN} \leq \overline{H}^{\reg}(Y \vert X)_{\mN}
\ee
where in the first equality we used the additivity of $\overline{H}$ as shown in Lemma \ref{lm:maxextadd}. Therefore, all the inequalities above must be equalities so that $\H(Y \vert X)_{\mN}=\underline{H}^{\reg}(Y \vert X)_{\mN}=\overline{H}(Y \vert X)_{\mN}=\min_{x\in[m]}H(\p_x)$.
\end{proof}


\end{document}